%% file: ANA-SUSY-2018-37-PAPER.tex
\documentclass[PAPER, cernpreprint, atlasdraft=false, USenglish, texmf, orcidlogo]{atlasdoc}
\usepackage[subfigure]{atlaspackage}
\usepackage{atlasbiblatex}
\usepackage{multirow}
\usepackage{atlasphysics}
\usepackage{graphics}
\graphicspath{{logos/}{figures/}}

\usepackage{ANA-SUSY-2018-37-PAPER-defs}

\newcommand{\mli}[1]{\mathit{#1}}
\AtlasTitle{A search for $R$-parity violating supersymmetric decays of the top squark to a $b$-jet and a lepton in $\sqrt{s}$ = 13 TeV $\mli{pp}$ collisions with the ATLAS detector}

\AtlasAbstract{%
A search is presented for direct pair production of the stop,
the supersymmetric partner of the top quark, in a decay through
an $R$-parity violating coupling to a charged lepton and a $b$-quark.
The dataset corresponds to
an integrated luminosity of 140 $\mathrm{fb}^{-1}$ of proton--proton collisions at
a center-of-mass energy of $\sqrt{s} = 13$~$\text{Te\kern -0.1em V}$
collected between 2015 and 2018
by the ATLAS detector at the LHC.
The final state has two charged leptons (electrons or muons) and two $b$-jets.
The results of the search are interpreted in the context of a Minimal Supersymmetric Standard Model with an additional $B-L$ gauge symmetry that is spontaneously broken.
No significant excess is observed over the
Standard Model background and exclusion limits on stop pair production are set at 95\% confidence level.
The corresponding lower limits on the stop mass for 100\% branching ratios to a $b$-quark and an electron, muon, or tau-lepton are 1.9 TeV, 1.8 TeV and 800 GeV, respectively, extending the reach of previous LHC searches.
}

\AtlasRefCode{ANA-SUSY-2018-37} %

\PreprintIdNumber{CERN-EP-2024-136}

\HepDataRecord{144719}

\AtlasJournalRef{ Phys.\ Rev.\ D. 110 (2024) 092004}
\AtlasDOI{10.1103/PhysRevD.110.092004}

\hypersetup{pdftitle={ATLAS document},pdfauthor={The ATLAS Collaboration}}

\begin{document}

\maketitle

\tableofcontents

\section{Introduction}
\label{sec:intro}
%

% The next lines are included from the .//texinclude/Introduction.tex input file
\label{secref:intro}

The extension of the Standard Model (SM) of particle physics with
supersymmetry
(SUSY)~\cite{Miyazawa:1966,Ramond:1971gb,Golfand:1971iw,Neveu:1971rx,Neveu:1971iv,Gervais:1971ji,Volkov:1973ix,Wess:1973kz,Wess:1974tw}
leads to processes that violate both baryon number ($B$)
and lepton number ($L$). This in turn may lead to rapid proton decay and
lepton-number-violating processes, such as decays of
$\mu \rightarrow e\gamma$, in conflict with experimental bounds.  A
conventional assumption to prevent these processes is to impose
conservation of a multiplicative quantum number
$R$-parity~\cite{Fayet:1976et,Fayet:1977yc,Farrar:1978xj,Fayet:1979sa,Dimopoulos:1981zb},
defined as $R=(-1)^{3(B-L)+2s}$ where $s$ is the spin of the particle.
$R$ has a value of $+1$ for SM particles and $-1$ for
SUSY particles.  $R$-parity conservation requires SUSY particles to be produced in pairs, and the
lightest supersymmetric particle (LSP) to be stable. The LSP
cannot carry electric charge or color charge without coming into
conflict with astrophysical data~\cite{Ellis, Pospelov}. At the LHC, the conventional
experimental signature for SUSY particles includes significant missing
transverse momentum due to the non-interaction of the LSP with the
detector.

An alternative approach is an $R$-parity violating (RPV) model which adds a local symmetry $U(1)_{B-L}$ to the
$SU(3)_C \times SU(2)_L \times U(1)_Y$ SM and includes three generations of
right-handed neutrino supermultiplets. The minimal supersymmetric extension then only
needs a vacuum expectation value (VEV) for a right-handed sneutrino in order
to spontaneously break the $B-L$ symmetry. The size of the RPV coupling is directly related to the right-handed sneutrino VEV, and therefore to the neutrino sector. The size of the coupling is kept small by the small values of the neutrino mass ~\cite{FileviezPerez:2008sx, Barger:2008wn, FileviezPerez:2009gr, Everett:2009vy, Evans:1986ada, Lukas:1998yy, Braun:2005ux, Braun:2005nv, Braun:2006ae, Ambroso:2009jd, Ambroso:2010pe, Ovrut:2012wg, Deen:2016vyh, Deen:2016zfr}.  This minimal $B-L$
model, with a scalar top squark (stop or $\tilde{t}$)  decaying into a quark and a lepton, violates lepton number but not baryon number. This model is consistent
with proton stability and the bounds on lepton number violation~\cite{bounds}. The most noticeable effect of this small coupling
related to the neutrino mass is that the LSP can now decay via RPV processes, and
can carry color and electric charge.  This leads to unique
signatures~\cite{FileviezPerez:2012mj, Perez:2013kla, Ovrut:2012wg,
Ovrut:2014rba, Ovrut:2015uea} that are disallowed in conventional
models with $R$-parity conservation.

A novel possibility with RPV is that the LSP could be a stop, where the large mass of the top quark acts to make the
stop significantly lighter than the other squarks due to
renormalization group effects~\cite{Barbieri:1987fn,
deCarlos:1993rbr, allanach2024supersymmetry}.  Although, according to theory predictions based on this model, a stop LSP is much less likely than
an electroweak gaugino LSP even when taking naturalness considerations
into account, it is interesting nonetheless to
perform a search for the stop in the LHC data. The stop pair
production rate via the strong interaction is higher than the pair
production rate of electroweak gauginos in the experimentally
accessible mass ranges~\cite{Ovrut:2014rba, Ovrut:2015uea}.
The total width of the stop is expected to be negligible relative to experimental resolution according to this model.
Theory predicts stop pair production cross-sections from $610$ fb to $0.015$ fb for 500~\GeV to 2~\TeV mass stops~\cite{Beenakker_2016}.

\begin{figure}[t]
\centering{
\includegraphics[scale=0.25]{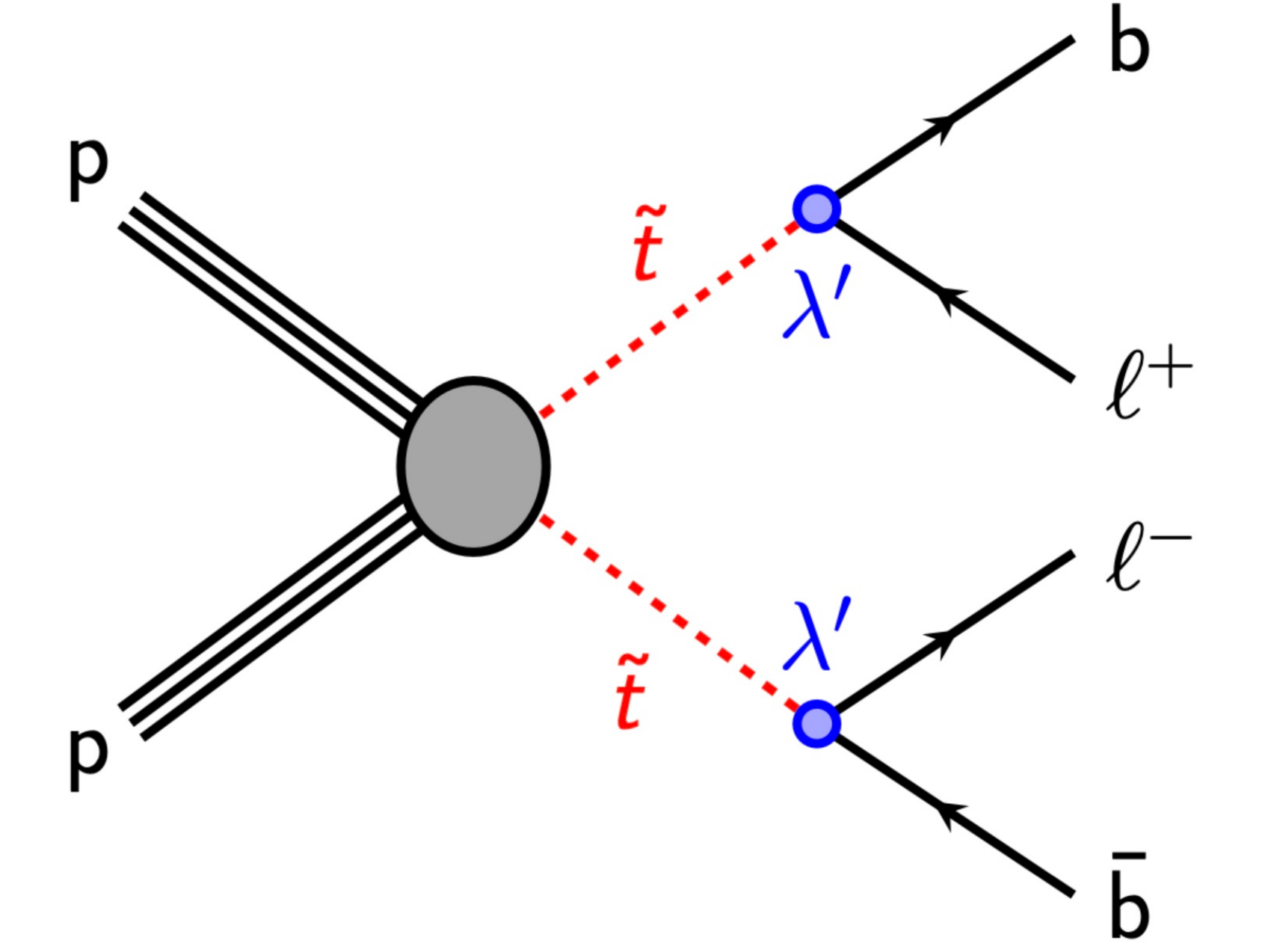}%
}
\caption{Diagram for stop pair production, with each stop
decaying into a charged lepton and a $b$-quark.}
\label{fig:feynman_diagram}
\end{figure}

This paper presents a search for direct stop pair production,
with the decay of each stop via an
RPV interaction to a charged lepton and a $b$-quark
(see Figure~\ref{fig:feynman_diagram}), based on 140 $\mathrm{fb}^{-1}$ of proton--proton collisions at
$\sqrt{s} = 13$~$\text{Te\kern -0.1em V}$ recorded with the ATLAS detector.
Previous related results include the RPV stop search with a partial Run 2 dataset with an integrated luminosity of 36.1 $\mathrm{fb}^{-1}$ \cite{SUSY-2016-29}, the RPV chargino/neutralino search with a trilepton resonance~\cite{SUSY-2018-36} and the search for top squarks in final states with two top quarks and several light-flavored jets~\cite{CMS-SUS-19-004}. Leptoquark results and RPV long-lived searches also include similar final states to this analysis \cite{EXOT-2019-13, leptoquark1, EXOT-2019-12, leptoquark3, EXOT-2019-19, leptoCMS, CMS-EXO-19-021}. In contrast to $R$-parity conserving
searches for the stop, there is no significant missing transverse
momentum. The stop decay branching ratios to $e b$, $\mu b$, and
$\tau b$ depend on the neutrino mass
hierarchy~\cite{Marshall:2014cwa, Marshall:2014kea}. The
experimental signature comprises of two oppositely charged leptons, which do not have to be the same flavor,
and two $b$-jets.  For this analysis, only events with
electron or muon signatures are selected, and final states are split
by flavor into \EE, \EM, and \MM channels. Sensitivity for a stop decaying to a $b$-jet and $\tau$-lepton (through leptonic $\tau$ decay) is obtained through investigation of these electron and muon channels. A loose requirement of one or more identified $b$-jets is used to maximize signal selection efficiency for high values of the stop mass.
There are two permutations of the leptons and jets to form lepton-jet pairs; the permutation with the smallest mass
asymmetry is chosen to reconstruct the stop decays correctly with high efficiency and the value of the asymmetry is
required to be small to reject background.
The invariant mass of each lepton-jet resonance is required to be above the top quark mass to further reduce background from top quark
pair and single top quark production. This analysis extends the reach of previous searches by performing a fit to
the distribution of the mass of the leading lepton-jet pair.

The ATLAS detector is described in \secref{detector}, with the dataset collected during Run 2 of the LHC and
the corresponding Monte Carlo simulation samples presented in \secref{mcsim}. The identification and
reconstruction of jets and leptons is presented in \secref{eventreco}, and the discriminating variables used to
construct the signal regions are described in \secref{sel}. The method of background estimation is
described in \secref{background}, and the systematic uncertainties are detailed in \secref{systs}. Results are presented in \secref{results}.

% End of text imported from the .//texinclude/Introduction.tex input file

\FloatBarrier
\section{ATLAS detector}
\label{sec:detector}
%

% The next lines are included from the .//texinclude/atlas-detector.tex input file
%
\newcommand{\AtlasCoordFootnote}{%
ATLAS uses a right-handed coordinate system with its origin at the nominal interaction point (IP)
in the centre of the detector and the \(z\)-axis along the beam pipe.
The \(x\)-axis points from the IP to the centre of the LHC ring,
and the \(y\)-axis points upwards.
Polar coordinates \((r,\phi)\) are used in the transverse plane,
\(\phi\) being the azimuthal angle around the \(z\)-axis.
The pseudorapidity is defined in terms of the polar angle \(\theta\) as \(\eta = -\ln \tan(\theta/2)\),
and is equal to the rapidity $ y = \frac{1}{2} \ln \left( \frac{E + p_z c}{E - p_z c} \right) $ in the relativistic limit.
Angular distance is measured in units of \(\Delta R \equiv \sqrt{(\Delta y)^{2} + (\Delta\phi)^{2}}\).}

\label{sec:detector}

The ATLAS detector~\cite{PERF-2007-01} at the LHC covers nearly the entire solid angle around the collision point.\footnote{\AtlasCoordFootnote}
It consists of an inner tracking detector surrounded by a thin superconducting solenoid, electromagnetic and hadron calorimeters,
and a muon spectrometer incorporating three large superconducting air-core toroidal magnets.

The inner-detector system (ID) is immersed in a \qty{2}{\tesla} axial magnetic field
and provides charged-particle tracking in the range \(|\eta| < 2.5\).
The high-granularity silicon pixel detector covers the vertex region and typically provides four measurements per track,
the first hit normally being in the insertable B-layer installed before Run~2~\cite{ATLAS-TDR-19,PIX-2018-001}.
It is followed by the silicon microstrip tracker, which usually provides eight measurements per track.
These silicon detectors are complemented by the transition radiation tracker (TRT),
which enables radially extended track reconstruction up to \(|\eta| = 2.0\).
The TRT also provides electron identification information
based on the fraction of hits (typically 30 in total) above a higher energy-deposit threshold corresponding to transition radiation.

The calorimeter system covers the pseudorapidity range \(|\eta| < 4.9\).
Within the region \(|\eta|< 3.2\), electromagnetic calorimetry is provided by barrel and
endcap high-granularity lead/liquid-argon (LAr) calorimeters,
with an additional thin LAr presampler covering \(|\eta| < 1.8\)
to correct for energy loss in material upstream of the calorimeters.
Hadron calorimetry is provided by the steel/scintillator-tile calorimeter,
segmented into three barrel structures within \(|\eta| < 1.7\), and two copper/LAr hadron endcap calorimeters.
The solid angle coverage is completed with forward copper/LAr and tungsten/LAr calorimeter modules
optimized for electromagnetic and hadronic energy measurements respectively.

The muon spectrometer (MS) comprises separate trigger and
high-precision tracking chambers measuring the deflection of muons in a magnetic field generated by the superconducting air-core toroidal magnets.
The field integral of the toroids ranges between \num{2.0} and \qty{6.0}{\tesla\metre}
across most of the detector.
Three layers of precision chambers, each consisting of layers of monitored drift tubes, cover the region \(|\eta| < 2.7\),
complemented by cathode-strip chambers in the forward region, where the background is highest.
The muon trigger system covers the range \(|\eta| < 2.4\) with resistive-plate chambers in the barrel, and thin-gap chambers in the endcap regions.

Events are selected by the first-level trigger system implemented in custom hardware,
followed by selections made by algorithms implemented in software in the high-level trigger~\cite{TRIG-2016-01}.
The first-level trigger accepts events from the \qty{40}{\MHz} bunch crossings at a rate below \qty{100}{\kHz},
which the high-level trigger further reduces in order to record complete events to disk at about \qty{1}{\kHz}.

A software suite~\cite{SOFT-2022-02} is used in data simulation, in the reconstruction
and analysis of real and simulated data, in detector operations, and in the trigger and data acquisition
systems of the experiment.

% End of text imported from the .//texinclude/atlas-detector.tex input file

\FloatBarrier
\section{Data and simulated event samples}
\label{sec:mcsim}
%

% The next lines are included from the .//texinclude/mcsim.tex input file
\label{secref:mcsim}

This analysis is performed using data from the LHC \pp collisions with center-of-mass energy of $\sqrt{s} = 13$~$\text{Te\kern -0.1em V}$, collected
during 2015 -- 2018 with the ATLAS detector. The total integrated luminosity of this dataset is 140 $\mathrm{fb}^{-1}$. The uncertainty of the combined 2015 -- 2018 integrated luminosity is 0.83\%~\cite{DAPR-2021-01}, obtained by the LUCID-2 detector~\cite{LUCID2} for the primary luminosity measurements, complemented by measurements using the inner detector and calorimeters. Due to the high instantaneous luminosity and large total inelastic \pp cross-section, there are, on average 33.7 simultaneous ("pileup") collisions in each bunch crossing. Data events must satisfy quality requirements to be included in the analysis~\cite{DAPR-2018-01}.

Monte Carlo (MC) simulation is used to predict the backgrounds from SM processes,
estimate the detector response and efficiency to reconstruct the signal process and the associated systematic uncertainties.
The largest sources of SM background are top quark pair production (\ttbar), single top quark production (single top), and \Zjet production,
and their yields are estimated through data-driven methods described in \secref{background}.
Smaller backgrounds originate from \Wjet, $t\bar{t} + V (V = W,Z)$, and triboson and diboson (vector boson) production and are estimated directly from MC simulation.
Details of the MC simulation are given below and are summarized in \tabref{mc}.

The production of \ttbar events was modeled using the
\POWHEGBOX[v2]~\cite{Nason:2004rx,Frixione:2007vw,Alioli:2010xd}
generator at next-to-leading order (NLO) in QCD with the \NNPDF[3.0nlo]~\cite{Ball:2014uwa} set of parton distribution functions (PDF)
and the \hdamp parameter\footnote{The
\hdamp parameter is a resummation damping factor and one of the
parameters that controls the matching of \POWHEG matrix elements to
the parton shower and thus effectively regulates the
high-\pT radiation against which the \ttbar system recoils.} set
to 1.5 times the mass of the top quark, which is about 172.5 \GeV~\cite{ATL-PHYS-PUB-2016-020}. The associated production of top quarks with \(W\) bosons (single top \(Wt\)) was
modeled by the
\POWHEGBOX[v2]%
generator at NLO in QCD using the five-flavor scheme and the
\NNPDF[3.0nlo] PDF set.~\cite{Ball:2014uwa}.
The diagram removal scheme~\cite{Frixione:2008yi} was used to
remove interference and overlap with \ttbar production.
The uncertainty on the single top yield due to the destructive interference between the \ttbar and \Wt processes is estimated by using inclusive $WWbb$ samples generated at leading order (LO) using
\MGNLO[2.3.3]~\cite{Alwall:2014hca}.
The events were interfaced to \PYTHIA[8.230]~\cite{Sjostrand:2014zea} for the modeling of the parton shower (PS), hadronization and the underlying event (UE), using the A14
set of tuned parameters (tune)~\cite{ATL-PHYS-PUB-2014-021} and the \NNPDF[2.3lo] set of
PDFs~\cite{Ball:2012cx}.

Due to the majority of \ttbar and single top events being produced with significantly lower \Ht than the targeted signal region,
special \Ht -sliced \ttbar and single top samples were utilized. \Ht is defined as the scalar sum of the \pt of the decay products.
These samples apply \Ht requirements on produced events in order to boost available statistics at high \Ht.
This improves estimation of systematic uncertainties on the background in the signal regions with the highest values of the leading lepton-jet mass.
Other than the additional \Ht filter, the generation configuration is identical to the nominal samples.

MC samples modeling pairs of stops decaying into \bl are generated for use in optimizing the selections
and estimating the analysis sensitivity.
Signal samples are generated at LO in QCD for stop masses between 600 \GeV and 1900 \GeV in increments of 50 \GeV using \MGNLO[2.6.2]~\cite{Alwall:2014hca} interfaced with \PYTHIA with the \CTEQ[6L1] PDF set~\cite{Alwall:2014hca,Pumplin:2002vw}. Signal cross-sections are calculated at NLO in QCD in the strong coupling constant, adding the resummation of soft-gluon emission at next-to-leading-logarithmic accuracy (NLO+NLL)~\cite{Beenakker:1997ut,Beenakker:2010nq,Beenakker:2011fu}. The nominal cross-sections and the uncertainty are taken from an envelope of cross-section predictions using different PDF sets and factorization and normalization scales, as described in Ref.~\cite{Borschensky:2014cia}.
The stops are allowed to decay directly into a $b$-quark and either an
electron, muon, or $\tau$-lepton, with identical branching ratios given by
$\mathcal{B}(\tilde{t}\rightarrow eb)
= \mathcal{B}(\tilde{t}\rightarrow \mu b)
= \mathcal{B}(\tilde{t}\rightarrow \tau b) = 1/3$.
In the interpretation of the search results, various hypotheses were tested for the branching ratios by reweighting events with appropriate factors so that the branching ratios for all three charged lepton flavors sum to unity.

The modeling of $c$-hadron and $b$-hadron decays was performed with \textsc{EvtGen}~1.6.0~\cite{Lange:2001uf}.
\textsc{EvtGen} is used for all samples except those generated with Sherpa. Events from all generators were propagated through a full simulation of the ATLAS detector using \GEANT~\cite{Agostinelli:2002hh} to model the interactions of particles with the detector. The generation of the simulated event samples includes the effect of multiple $\mli{pp}$ interactions per bunch crossing, as well as the effect on the detector response due to interactions from bunch crossings before or after the one containing the hard interaction.

\begin{table}[t]
\footnotesize
\centering{
\label{tab:mc}
\caption{MC simulation details by physics process.}
\resizebox{\columnwidth}{!}{%
\begin{tabular}{lccccc}
\toprule
Process & Event & PS and & UE tune & PDF & Cross-section order\\
& generator & hadronization & & & \\
\hline
\hline

\ttbar   & \POWHEGBOX~v2 & \PYTHIA[8] & A14 & \NNPDF[3.0nlo]  & NNLO+NNLL ~\cite{Frixione:2007nw,Nason:2004rx,Frixione:2007vw,Alioli:2010xd}\\
single top (\Wt) & \POWHEGBOX~v2 & \PYTHIA[8] & A14 & \NNPDF[3.0nlo] & NNLO+NNLL~\cite{Re:2010bp,Nason:2004rx,Frixione:2007vw,Alioli:2010xd}\\
$W/Z$+jets &  \Sherpa 2.2.1  & \Sherpa 2.2.1 & Default & NNPDF3 & LO+NLO~\cite{Bothmann:2019yzt,Ball:2014uwa,Anastasiou:2003ds}\\
Diboson,triboson  & \Sherpa 2.2.2 & \Sherpa 2.2.2 & Default & NNPDF3 & LO+NLO~\cite{Bothmann:2019yzt,Gleisberg:2008fv,Schumann:2007mg} \\
\ttbar+ $V$  & \MGNLO[2.3.3] & \PYTHIA[8]  & A14 & NNPDF3 & NLO~\cite{Alwall:2014hca} \\
$\tilde{t}\bar{\tilde{t}}$ signal & \MGNLO[2.6.2] & \PYTHIA[8] & A14 & \CTEQ[6L1] & NLO+NLL~\cite{Beenakker:1997ut,Beenakker:2010nq,Beenakker:2011fu} \\
\bottomrule
\end{tabular}}
\label{tab:mc}}
\end{table}

% End of text imported from the .//texinclude/mcsim.tex input file

\FloatBarrier

\section{Physics object reconstruction}
\label{sec:eventreco}
%

% The next lines are included from the .//texinclude/eventreconstruction.tex input file
\label{sec:reco}
Each event is required to have a primary reconstructed vertex with two or more associated tracks,
where the primary vertex is chosen as the vertex with the highest $\Sigma \pt^{2}$ of associated tracks~\cite{ATL-PHYS-PUB-2015-026}.
Two stages of quality and kinematic requirements are applied to leptons and jets.
The looser baseline requirements are applied first, and baseline leptons and jets are used to resolve
any misidentification or overlap among electrons, muons, and jets.
The subsequent tighter signal requirements are then applied to identify high-quality leptons and jets
in the kinematic phase space of interest.

Electron candidates are reconstructed from energy deposits in the
electromagnetic calorimeter matched to a charged-particle track in the ID.
Baseline electron candidates must have $\pt > 10~\GeV$, $|\eta| < 2.47$, and satisfy a baseline \textit{Loose}
electron likelihood identification~\cite{EGAM-2018-01}.
Each baseline electron must have a longitudinal impact parameter with respect to the primary vertex ($z_0^{\mathrm{PV}}$)
that satisfies $|z_0^{\mathrm{PV}}~\text{sin}\theta| < 0.5$~mm.
Signal electrons must pass the baseline electron selection, have $\pt > 40~\GeV$,
and satisfy a \textit{Tight} electron likelihood identification.
In addition, they must be isolated from nearby activity, satisfying a loose \pt-dependent
track-based criterion~\cite{EGAM-2018-01}.
Finally, their trajectory must be consistent with the primary vertex, such that their impact parameter
in the transverse plane ($d_0^{\mathrm{PV}}$) satisfies $|d_0^{\mathrm{PV}}|/\sigma_{d_0^{\mathrm{PV}}} < 5$, where $\sigma_{d_0^{\mathrm{PV}}}$
is the uncertainty in $d_0^{\mathrm{PV}}$.
Muon candidates are reconstructed by combining tracks in the ID with tracks in the MS.
Baseline muon candidates must have \pt $>10~\GeV$, $|\eta| < 2.7$, $|z_0^{\mathrm{PV}}~\text{sin}\theta| < 0.5$~mm, and satisfy the
\textit{Medium} muon identification criteria~\cite{MUON-2018-03}.
Signal muons must pass the baseline muon selection, have $\pt > 40~\GeV$, $|\eta| < 2.7$ and $|d_0^{\mathrm{PV}}|/\sigma_{d_0^{\mathrm{PV}}} < 3$.
As with electrons, muons must satisfy the \pt-dependent loose track-based isolation criteria~\cite{MUON-2018-03}.
Events containing a poorly measured signal muon, as determined by having
incompatible momentum measurements in the ID and the MS, are rejected.
Absolute requirements of $|z_0^{\mathrm{PV}}| < 1$~mm and $|d_0^{\mathrm{PV}}| < 0.2$~mm on the impact parameters of signal muons are applied to reject cosmic muons.

Reconstructed jets are identified by combining measurements from both the ID and the calorimeter using a particle flow (PFlow) algorithm.
PFlow objects defined by this algorithm are then passed as inputs to the anti-$k_{t}$ algorithm~\cite{PERF-2015-09,Cacciari:2011ma} with a radius
parameter $R = 0.4$.
Jets are further calibrated to account for the predicted detector response in MC simulation,
and a residual calibration of jets in data is derived through \textit{in situ} measurements~\cite{JETM-2018-05}.
Baseline jet candidates are required to have $\pt >20~\GeV$ and $|\eta| < 4.5$.
Jets with $\pt< 60~\GeV$ and $|\eta| < 2.4$ must satisfy requirements on the jet vertex tagger~\cite{Aad:2015ina}, which is used to reject pileup jets.
Signal jets must pass the baseline jet selection and have $\pt > 60~\GeV$.
Events are rejected if they contain a jet that fails the \textit{Loose} quality criteria~\cite{ATLAS-CONF-2015-029}, reducing contamination from noise bursts and non-collision backgrounds.

The identification of baseline jets containing $b$-hadrons ($b$-jets) is performed using the DL1r $b$-tagging algorithm~\cite{DL1d}.
The $b$-tagging requirements result in an efficiency of 85\% for jets containing $b$-hadrons, as determined in a sample of simulated \ttbar events with final states containing two leptons, and the corrections are consistent with unity with uncertainties at the level of a few percent over most of the jet \pT range.
The 85\% working point has a rejection factor of 3.1 and of 33 on charm and light-jets, respectively~\cite{FTAG-2018-01}.

To avoid reconstructing a single detector signature in multiple ways, an overlap removal
procedure is performed on baseline leptons and jets.
The requirements are applied sequentially, and failing particles are removed from consideration in the subsequent steps.
If an electron and muon share a track in the ID, the electron is removed.
Any jet that is not $b$-tagged and is within a distance
$\Delta R(e,\mathrm{jet}) \le 0.2$ of an electron is removed.
Any electron with $\Delta R(e,\mathrm{jet}) \le 0.4$ from a baseline jet is removed since this electron is assumed to be a constituent of the jet.
Any jet that is within $\Delta R(\mu,\mathrm{jet}) \le 0.2$ and has less than three tracks is removed. If that jet is $b$-tagged, the muon is instead removed.
Any muon with $\Delta R(\mu, \mathrm{jet}) \le 0.4$ of a baseline jet is then removed.

The trigger, reconstruction, identification, and isolation efficiencies of electrons~\cite{EGAM-2018-01} and muons~\cite{MUON-2018-03}
in MC simulation are corrected using events in data with leptonic $Z$ and $\Jpsi$ decays.
Similarly, corrections to the $b$-tagging efficiency and mis-tag rate in MC simulation are derived from various control regions in data~\cite{FTAG-2018-01,FTAG-2019-02,FTAG-2020-08}.

% End of text imported from the .//texinclude/eventreconstruction.tex input file

\FloatBarrier
\section{Event selection}
\label{sec:sel}
%

% The next lines are included from the .//texinclude/eventselection.tex input file
\label{sec:sel}
To identify the pair production of stops, events are required to have at least two leptons and two jets.
If more than two leptons or two jets are found, the two highest-\pt leptons and jets are selected.
At least one of the two leading jets must be $b$-tagged.
The selected leptons are required to have opposite charge, and one of them must be consistent with the associated single-lepton trigger~\cite{TRIG-2018-05,TRIG-2018-01}.
The lepton is not required to be the leading lepton in the event, but must exceed the offline \pT threshold for that trigger. The offline \pT requirement for each trigger is set to $1~\GeV$ above online for electron triggers and 5\% above for muon triggers. The trigger requirement is applied in both data and MC. To account for the difference in trigger efficiency, a trigger scale factor is applied to MC events that passed the trigger requirement~\cite{trigger}. This trigger requirement is highly efficient for signal events, with an efficiency of 93\% for the $\mu\mu$ channel and
98\% for the $ee$ channel.

The lepton-jet pair from each stop decay reconstructs the invariant mass $\Mbl$ of the original stop.
In an event with two leptons and two jets, two pairings are possible; one that reconstructs the correct
stop masses, and one which inverts the pairing and incorrectly reconstructs the masses.
As the two masses for the correct pairing should be roughly equal, the pairing that minimizes the mass asymmetry between $\Mbl^0$ and $\Mbl^1$ is chosen, defined as

\begin{equation*}
\MblAsym = \frac{\Mbl^0-\Mbl^1}{\Mbl^0+\Mbl^1}.
\label{eq:asym}
\end{equation*}

Here $\Mbl^0$ is chosen to be the larger of the two masses.
Events are further selected to have small mass asymmetry $\MblAsym < 0.2$.
This reduces the contamination from background processes with a more uniform $\MblAsym$ distribution.
The power of the asymmetry variable can be seen in the reference to the previous iteration of this search.~\cite{SUSY-2016-29}
Several other kinematic selections are defined to reduce the contribution from the largest backgrounds.
As the stop decay products are generally very energetic, a selection on their scalar \pt sum is applied such that $\Ht > 1000~\GeV$, with \Ht defined as the scalar sum of the \pT of the selected two leptons and two jets.
To reduce contamination from \Zjets events, a requirement is placed on the invariant mass of two leptons of $\mll > 300~\GeV$.
A large fraction of the background from processes involving a top quark is suppressed through the requirement on $\Mbl^0$ and $\MblAsym$,
with correctly paired top quark masses falling well below the signal region (SR) requirements.
However, top quark decays in which the lepton and $b$-jet decay products are mispaired can enter the SRs if the invariant masses happen to be large.
In such cases it is the rejected pairing that properly reconstructs the decay of a top quark, with one of the two \bl pair masses
below the kinematic limit for a top quark decay.
To suppress such backgrounds, events are rejected if the subleading \bl mass of the rejected pairing, $m_{b\ell}^{1, \text{rej}}$,
is indicative of that of a reconstructed top quark, with $m_{b\ell}^{1 \text{,rej}} < 150~\GeV$. %

All of the SRs require the leading lepton-jet mass to be above 400~\GeV. A set of 15 variable-width bins in the leading lepton-jet mass were chosen from
optimization studies based on simulated signal and backgrounds. The signal width of the reconstructed lepton-jet resonances increases with mass, from $\sim$ 20 -- 80~\GeV for stop mass values of 600 -- 1900~\GeV. The optimization studies for the bin widths included the systematic uncertainty on the background separately for each bin.
In contrast to the previous search with two nested SRs above 800~\GeV and above 1100~\GeV~\cite{SUSY-2016-29}, this search extends the sensitivity to lower and higher values for the stop mass
and to lower values for the stop decay branching ratio to an electron and a $b$-jet or to a muon and a $b$-jet. A full list of region selections is given in \tabref{regions}.

%
%
%
%
%
%
%
%
%
%
%
%
%
%
%
%
%
%
%
%
%
%
%

%
%
%
%
%
%
%
%
%
%
%
%
%
%

% End of text imported from the .//texinclude/eventselection.tex input file

\FloatBarrier
\section{Background estimation}
\label{sec:background}
%

% The next lines are included from the .//texinclude/bkg_estimation.tex input file
\label{sec:background}
For each of the relevant backgrounds in the signal regions, one of two
methods is used to estimate the contribution.
The small \Wjet, \ttV, diboson and triboson backgrounds are estimated directly from MC simulation
and the normalization is corrected to the highest-order theoretical cross-section available.
For the dominant \ttbar, single top, and \Zjets backgrounds, the expected yield in the signal regions (SRs)
is estimated by scaling each MC prediction by a normalization factor derived from three dedicated control regions (CRs), one for each
background process. Each CR is defined to be
kinematically close to the SRs while inverting or relaxing specific
selections. This enhances the contribution from the targeted background
process in the CR designed for each major background while reducing the contamination from other backgrounds and
the benchmark signals.

To derive a background-only estimate, the normalizations of
the \ttbar, single top, and \Zjets backgrounds are determined through a
likelihood fit~\cite{Baak:2014wma} performed simultaneously to the
observed number of events in each CR.  The expected yield in each
region is given by the inclusive sum over all background processes in
the \EE, \EM, and \MM channels.  The normalization factor for each of
the \ttbar, single top, and \Zjets backgrounds are free parameters of the fit. The other systematic
uncertainties are treated as nuisance parameters in the fit and are not significantly constrained.

Several validation regions (VRs) are defined to test the extrapolation from the CRs to SRs over the relevant kinematic variables.
The VRs are disjoint from both the CRs and SRs, and are constructed to fall between one or more CRs and the SRs
in one of the extrapolated variables.
The VRs are not included in the fit, but provide a statistically independent cross-check of the background prediction.
Four VRs are constructed to test the extrapolation in the $\mbl^0$, $m_{b\ell}^{1 \text{,rej}}$, \mll and \Ht observables. Validation regions show signal contamination generally below 10\% for stop masses > 600~\GeV.
Details of the selection criteria in each CR and VR are provided in \tabref{regions}.

\begin{table}[t]
\begin{center}
\caption{Summary of the requirements used in signal, control, and validation regions. All regions require at least two jets and two oppositely charged leptons. In the CRZ both leptons are required to be of the same flavor.} %
\label{tab:regions}
\resizebox{\columnwidth}{!}{%
\begin{tabular}{c|ccccccc}
\toprule
Region &
$N_{b}$ &
$m_{b\ell}^0$  [GeV] &
$m_{b\ell}^{1 \text{,rej}}$ [GeV] &
$H_T$ [GeV] &
\MblAsym &
$m_{ll}$ [GeV] &
$m_{b\ell}^{0 \text{,rej}}$ [GeV] \\
\midrule
SR   & $\ge 1$ & $> 400$  & $> 150$ & $> 1000$   & $<0.2$ & >300 GeV & --  \\
\midrule
CRtt   & $\ge 1$ & [$180$, $500$]  & $<150$  & [$500$, $800$] & $< 0.2$ & >200 GeV & $< 180$ \\
CRst   & $=2$ & [$180$, $500$]  & $<150$  & [$400$, $800$] & $< 0.2$ & >200 GeV & $> 180$  \\
CRZ   & $\ge 1$ &  $>700$  & --  & $>1000$ & $<0.2$ & [$76.2$, $106.2$] & -- \\
\midrule
VR $m_{b\ell}^0$ & $\ge 1$ & $>500$ & $<150$  & [$600$, $800$] & $< 0.2$ & >300 GeV & -- \\
VR $m_{b\ell}^{1 \text{,rej}}$ & $\ge 1$ & [$200$, $500$]  & $>150$  & [$600$, $800$] & $< 0.2$ & >300 GeV & -- \\
VR $H_\textrm{T}$ & $\ge 1$ & [$200$, $500$]  & $<150$  & $>800$ & $< 0.2$ & >300 GeV & -- \\
VRZ & $=0$ & [$500$, $800$] & $>150$ & $>1000$ & $<0.2$ & >300 GeV & -- \\
\bottomrule
\end{tabular}}
\end{center}
\end{table}
\FloatBarrier

\subsection{Control regions}
\begin{figure}[t]
\centering
\subfigure[]{%
\includegraphics[width=0.47\textwidth]{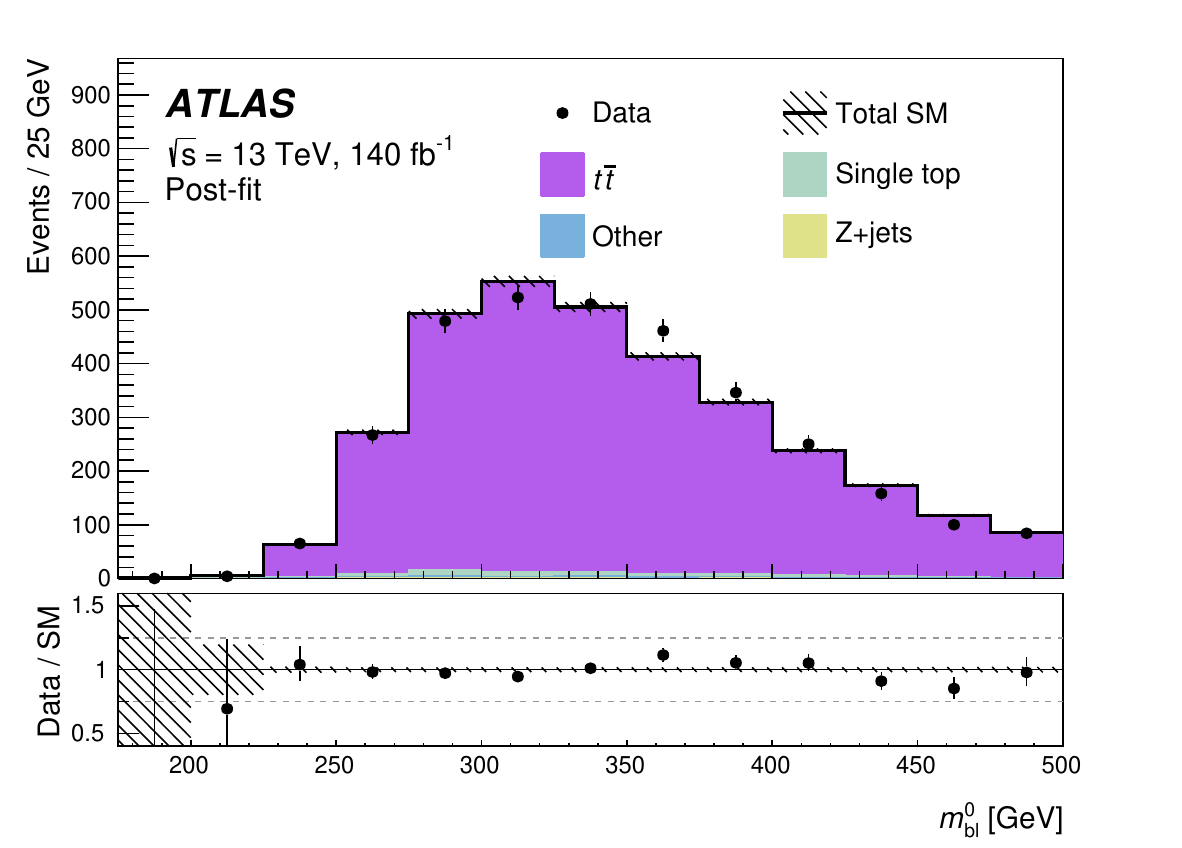}
\label{fig:CR_distributions:asym}
}
\subfigure[]{%
\includegraphics[width=0.47\textwidth]{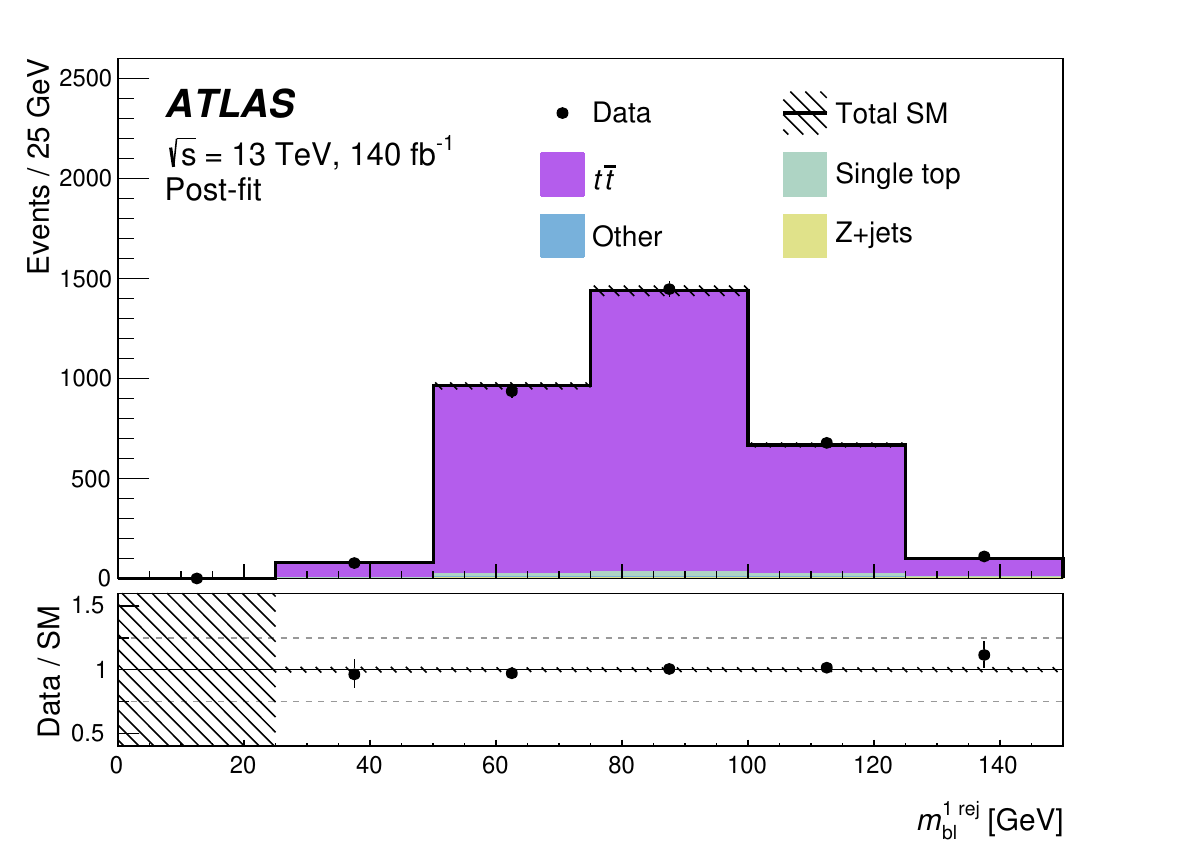}
\label{fig:CR_distributions:mll}
}
\subfigure[]{%
\includegraphics[width=0.47\textwidth]{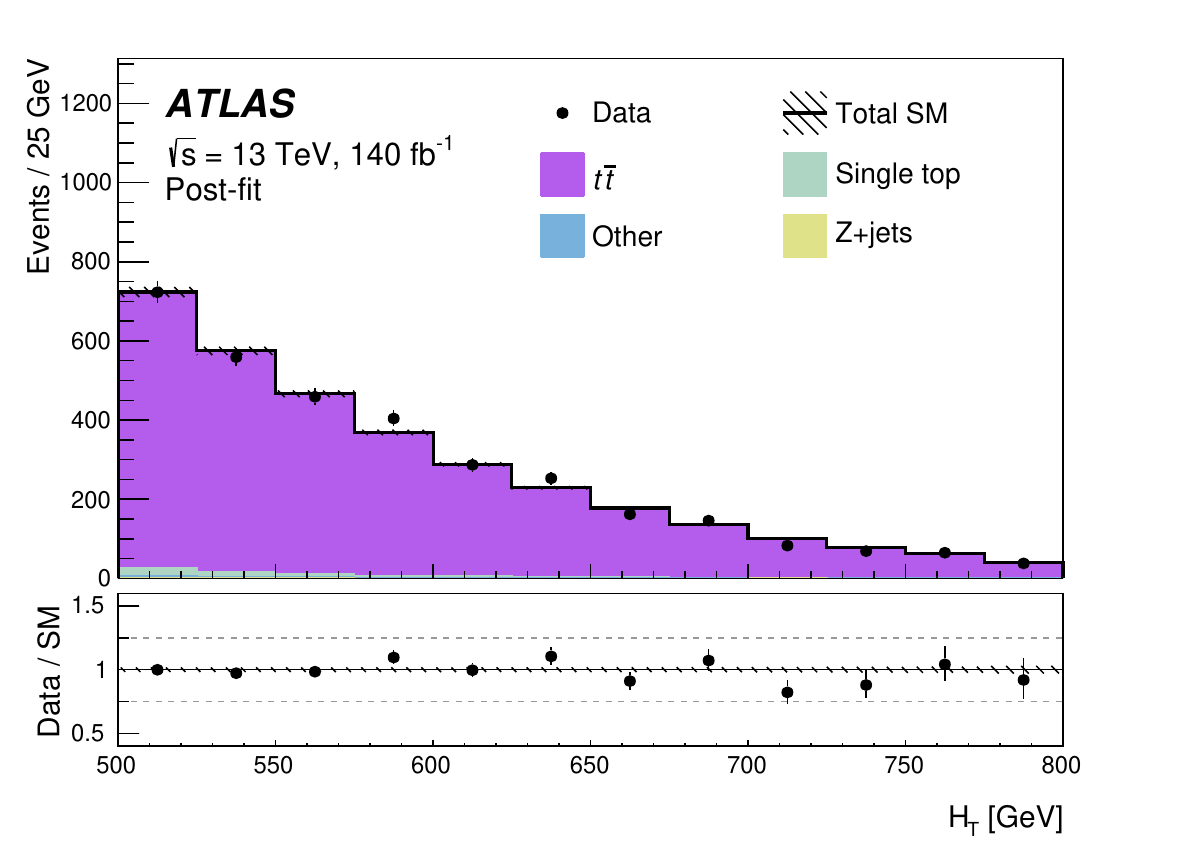}
\label{fig:CR_distributions:mblrej}
}
\subfigure[]{%
\includegraphics[width=0.47\textwidth]{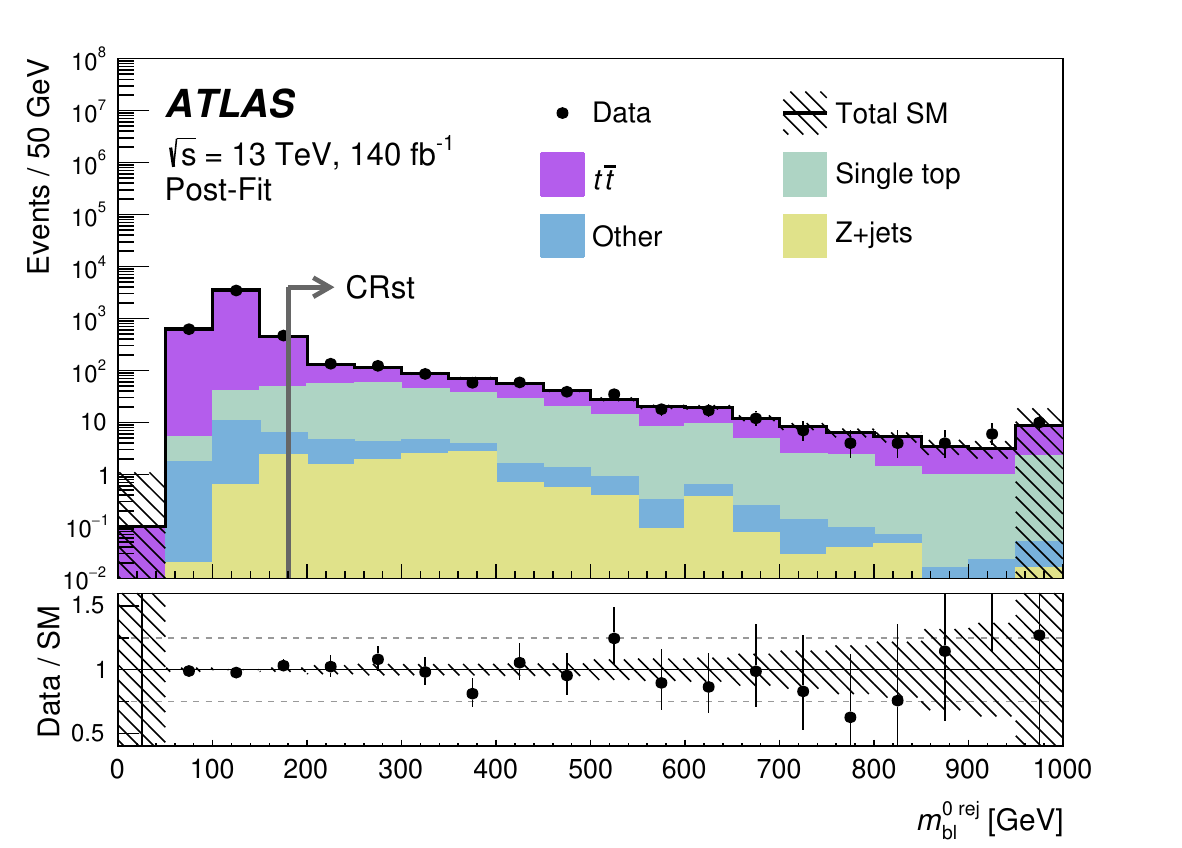}
\label{fig:CR_distributions:fourth}
}
\caption{
Distributions of \subref{fig:CR_distributions:asym} $m_{b\ell}^{0}$ in the CRtt, \subref{fig:CR_distributions:mll} $m_{b\ell}^{1, \text{rej}}$ in the CRtt, \subref{fig:CR_distributions:mblrej} \Ht in the CRtt, and \subref{fig:CR_distributions:fourth}  $m_{b\ell}^{0, \text{rej}}$ in the CRst for the data and post-fit MC prediction. Normalization factors are derived from the background-only fit configuration are applied to the dominant
\ttbar, single top, and \Zjet processes. The "Other" category consists of \Wjets, \ttV, diboson, and triboson processes. Processes are listed from largest to smallest contribution to the given region reading left to right and then top to bottom.
The relevant CR event selections are applied for each flavor inclusive distribution except the selection on the variable shown for \subref{fig:CR_distributions:fourth}. The arrow indicates the CR selection on the $m_{b\ell}^{0, \text{rej}}$ variable. The hatched band represents statistical and theoretical uncertainties on the backgrounds. The bottom panel of each plot shows the ratio between the data and the post-fit MC prediction. The last bin includes the overflow events.
}
\label{fig:CR_distributions}
\end{figure}

The \ttbar CR (CRtt) has specific requirements to separate it from the SR.
The SR selection on $\Mbl^{1 \text{,rej}}$ is inverted to $\Mbl^{1 \text{,rej}}<150$~\GeV. This enhances the fraction of \ttbar events where the jets and leptons are mispaired.
The average \Ht value needs to be lower in CRtt with respect to the SRs to remove contamination from signal events: a selection of $500 \le \Ht \le 800~\GeV$ is applied. The selection on \mll is loosened to \mll $>200$~\GeV and the $\Mbl^0$ selection is restricted to between 180 and 500~\GeV to enhance the \ttbar contribution in the CRtt. This \ttbar contribution comes from the incorrectly paired $\Mbl$. The $b$-tagging requirement is kept the same as the SR. %
The expected distributions of $\Mbl^{0}$, $\Mbl^{1, \text{rej}}$, and $H_{\textrm{T}}$
are shown in \figref{CR_distributions:asym}, \figref{CR_distributions:mll}, and \figref{CR_distributions:mblrej} for the \ttbar control region.

The \Zjet control region (CRZ) places selections on the dilepton pair to isolate the leptons coming from a $Z$ boson.
The dilepton invariant mass must be consistent with the mass of the $Z$ boson ($|\Mll - m_{Z}| \le 15$~\GeV) and the leptons are required to be of the same flavor, while still having opposite charge.
The $\Mll$ selection is very effective at removing signal contamination, which makes it possible to keep the \Ht and $\mbl^0$ of the CRZ events close to the SR. The requirements of $\Ht > 1000$~\GeV and $\Mbl^{0}>700~\GeV$ are placed.
The \MblAsym\ is required to be less than 0.2.

As single top production has a smaller cross-section than \ttbar, there is some difficulty with \ttbar contamination in the single top CR (CRst). To allow the single top background to have lower purity, the \ttbar CR was designed to have higher statistics to  constrain the \ttbar normalization factor. The purity for single top is about 30\% in CRst. The main difference between the CRst and CRtt is that the selection on $\Mbl^{0 \text{,rej}}$ is inverted. While \ttbar events are expected to have both $b\ell$ masses in the rejected pairing compatible with the top quark mass, only one is for single top backgrounds. This distinction allows for selecting the $\Mbl^{0 \text{,rej}}$ above the top quark mass to reject \ttbar events in the CRst. This region additionally requires both leading jets to be $b$-tagged, as this strengthens the discriminating power of the higher $\Mbl^{0 \text{,rej}}$ requirement. The expected distribution of $m_{bl}^{0, \text{rej}}$ is shown in \figref{CR_distributions:fourth} for the CRst. In the case of \ttbar events, requiring 2 $b$-tags improves the confidence that the correct jets are selected from the top quark decay, leading to at least one of the potential jet-lepton pairing schemes to properly reconstruct both top quarks. The selections $m_{b\ell}^{0} \ge 180$~\GeV and  $m_{b\ell}^{0 \text{,rej}} \ge 180$~\GeV are thus very effective at removing \ttbar events. The values of \Ht, \mll and $m_{b\ell}^{0}$ of the leading $\mathrm{b}\ell$ pair are constrained to reduce signal contamination, as described in \tabref{regions}.

\subsection{Validation regions}
Four disjoint validation regions are used to test the extrapolation of the background fit from the CRs to the SRs.
A full list of the region selections is given in \tabref{regions}.
The main extrapolations are from the \ttbar and single top CRs to the SR: in particular,
there is a significant extrapolation on the invariant mass of the leading accepted \bl pair $m_{bl}^0$, on the
invariant mass of the subleading rejected $b\ell$ pair $m_{b\ell}^{1 \text{,rej}}$, and on the \HT. For this reason, three separate
validation regions are defined by keeping all the \ttbar CR selections, and reverting the selection on these
three variables respectively. The upper bound on $\Mbl^{0 \text{,rej}}$ is also removed from the VRs.
Such regions allow for validation of the extrapolations from the CRs to the SRs.
The CRZ applies similar selections as the SR, with the exception of the $m_{\ell\ell}$ selection that needs to be modified
to select the $Z$-boson peak. For this reason, a VR is defined with exactly zero $b$-tagged jets, to enhance
the \Zjet background with respect to \ttbar and single top. The VR$m_{b\ell}^{1, \textrm{rej}}$ is divided into opposite-flavor leptons and same-flavor leptons, where the latter region allows a check of the estimate of the \Zjets background in the region with one or more $b$-tagged jets. All the other SR selections are applied, including $m_{\ell\ell}$
> 300~\GeV, with the only exception being $\Mbl^0$, for which a window of
[500, 800]~\GeV is used to reduce signal contamination.

\begin{figure}[t]
\centering
\includegraphics[width=0.85\textwidth]{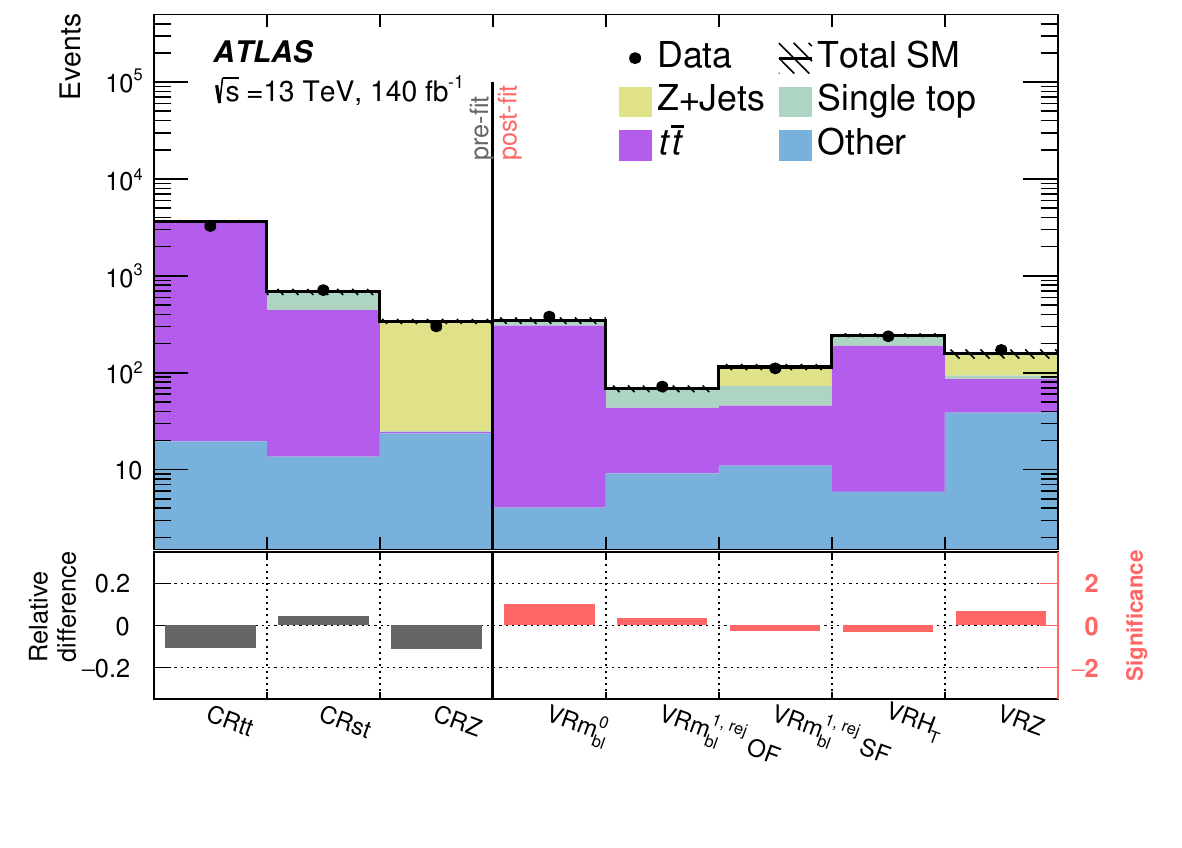}
\caption{
Comparison of the observed data and expected numbers of events in the CRs (pre-fit) and VRs (post-fit).
The "Other" category consists of \Wjets, \ttV, diboson and triboson processes. Processes are listed from largest to smallest contribution for the given region reading left to right and then top to bottom. The background prediction is derived with the background-only fit configuration.
The bottom panel of the plot shows the relative difference, the difference between the observed data and total background divided by the observed data, for the CRs as well as the significance of the difference for the VRs. The significance is computed through a profile likelihood method~\cite{Cousins_2008}.
}
\label{fig:regionYields}
\end{figure}

\begin{figure}[t]
\centering
\subfigure[]{%
\includegraphics[width=0.47\textwidth]{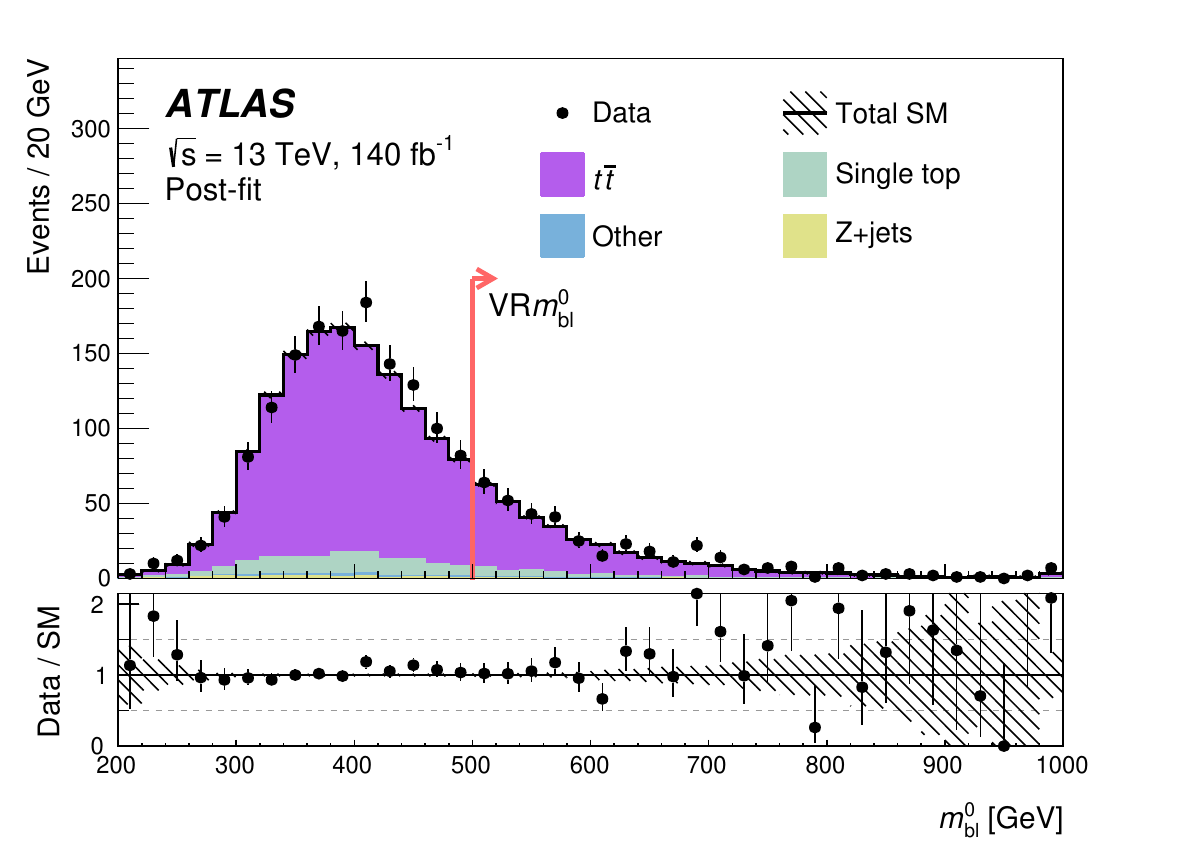}
\label{fig:VR_distributions:mbl0}
}
\subfigure[]{%
\includegraphics[width=0.47\textwidth]{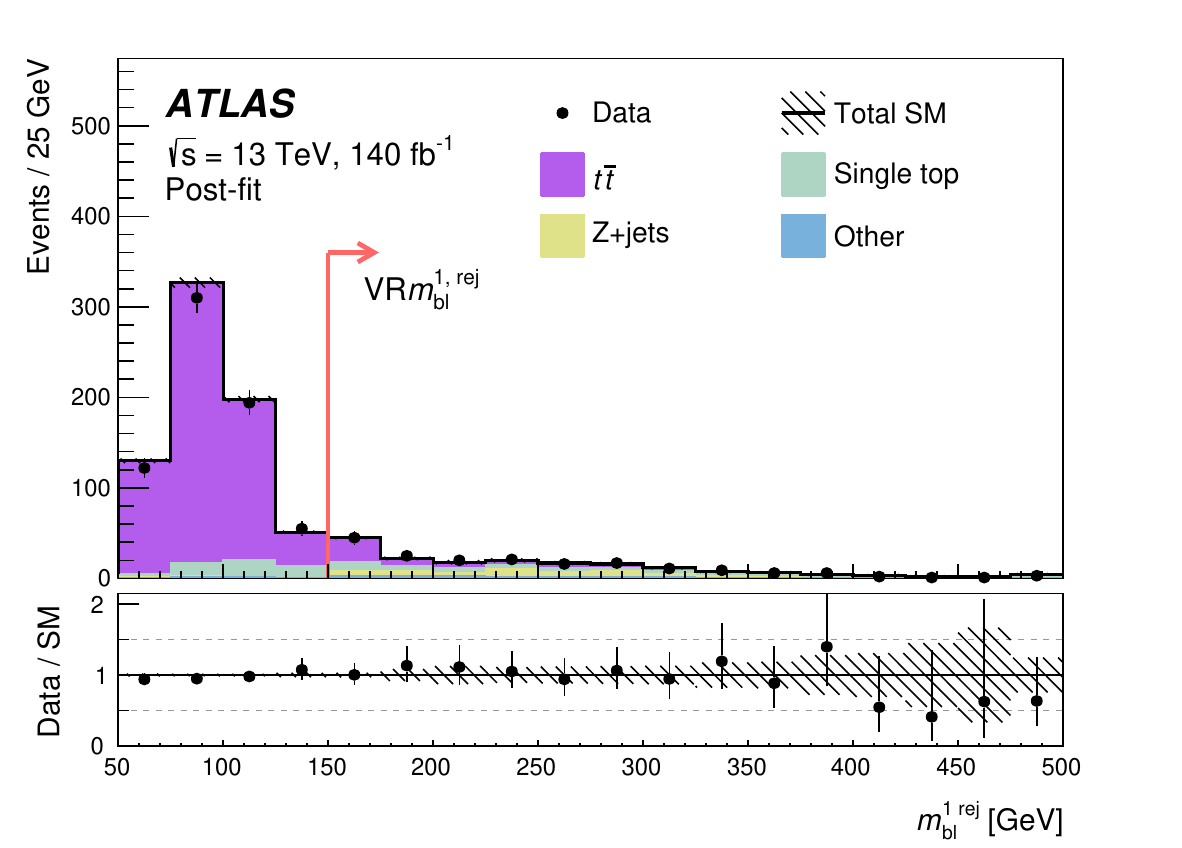}
\label{fig:VR_distributions:mblrej}
}
\subfigure[]{%
\includegraphics[width=0.47\textwidth]{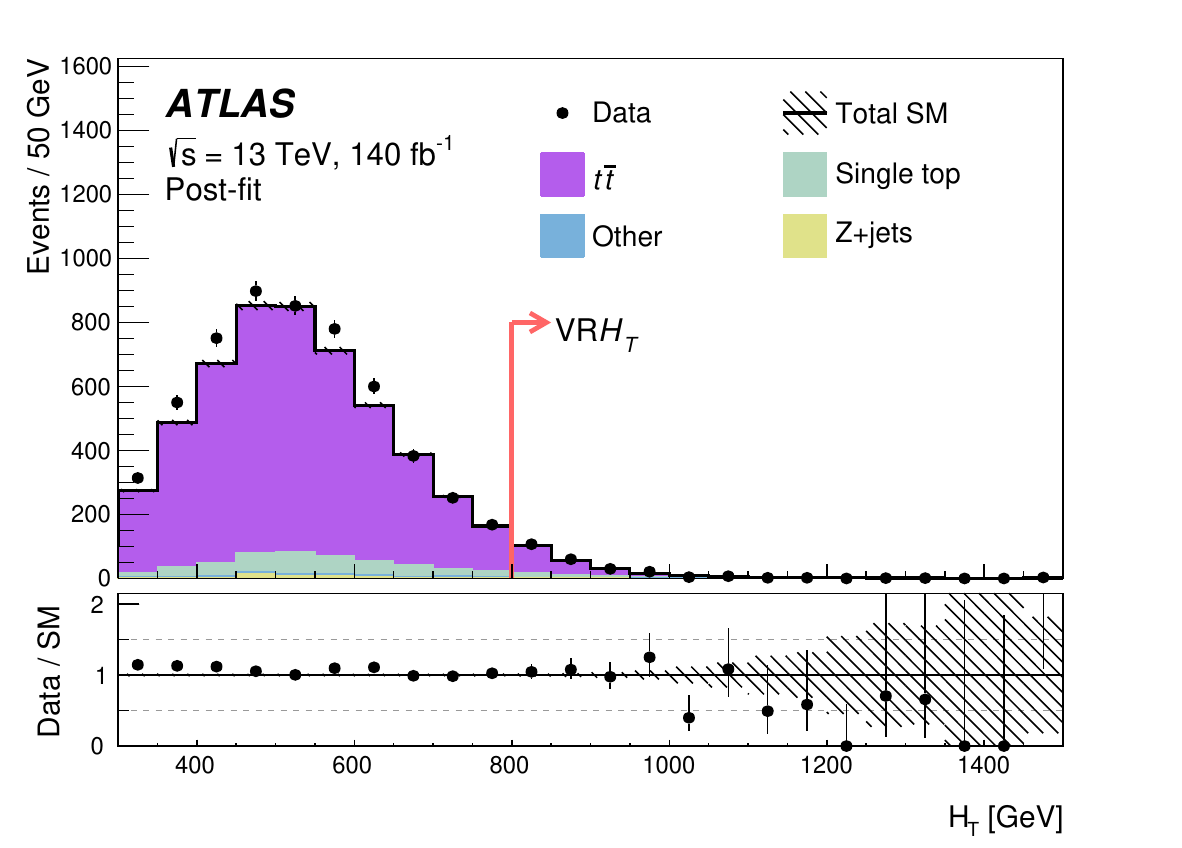}
\label{fig:VR_distributions:ht}
}
\subfigure[]{%
\includegraphics[width=0.47\textwidth]{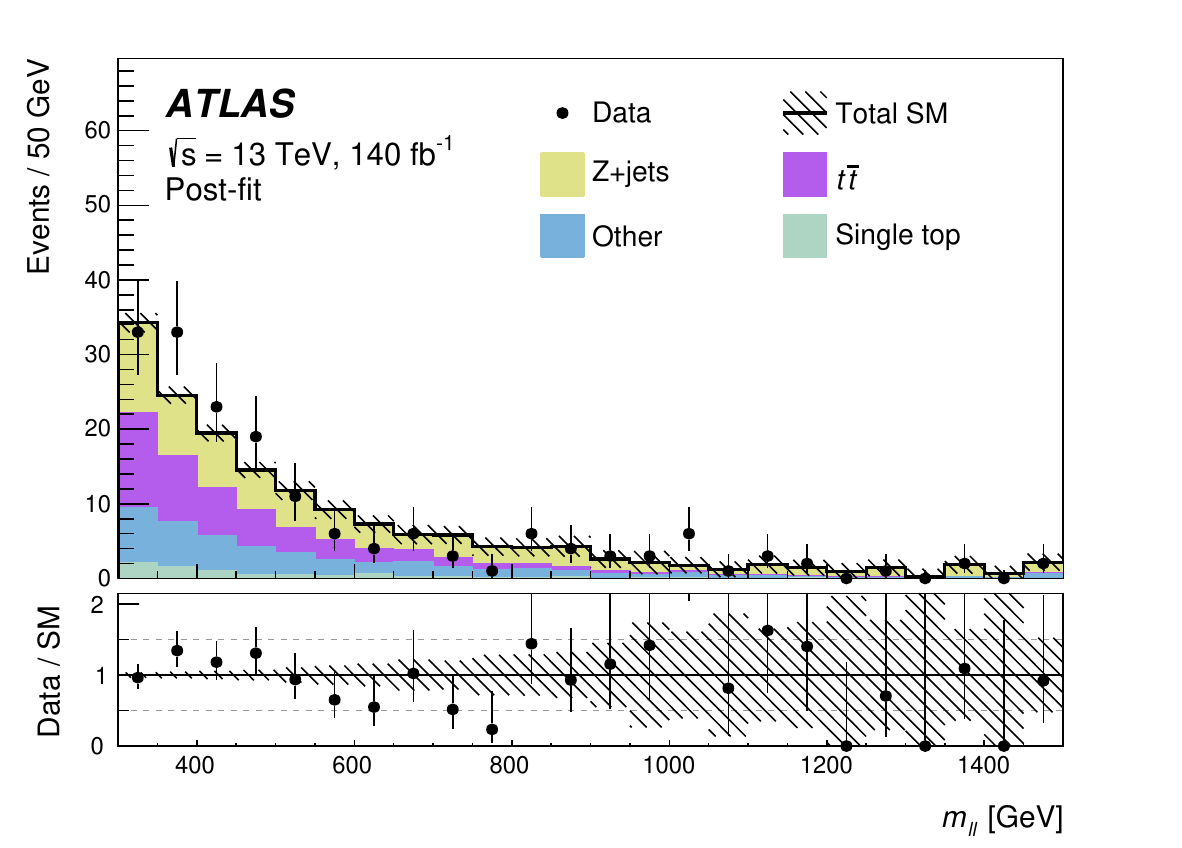}
\label{fig:VR_distributions:mll}
}
\caption{
Flavor-inclusive distributions of \subref{fig:VR_distributions:mbl0} \Mblacc in the \VRacc, \subref{fig:VR_distributions:mblrej} $\Mbl^{1 \text{,rej}}$ in the VR$m_{b\ell}^{1, \textrm{rej}}$,
\subref{fig:VR_distributions:ht} \Ht in the \VRht, and \subref{fig:VR_distributions:mll} \mll in the \VRZ for the data and post-fit MC prediction.
Normalization factors are derived from the background-only fit configuration and are applied to the dominant
\ttbar, single top, and \Zjets processes. The "Other" category consists of \Wjets, \ttV, diboson and triboson processes. Processes are listed from largest to smallest contribution to the given region reading left to right and then top to bottom. The relevant VR event selections are applied for each flavor-inclusive distribution except the selection on the variable shown for \subref{fig:VR_distributions:mbl0} \Mblacc in the \VRacc, \subref{fig:VR_distributions:mblrej} $\Mbl^{1 \text{,rej}}$ in the VR$m_{b\ell}^{1, \textrm{rej}}$,
and \subref{fig:VR_distributions:ht} \Ht in the \VRht. Figure 4\subref{fig:VR_distributions:mll} includes all relevant VR event selections. Regions outside of the proposed cut may show slight discrepancy from applying normalization factors designed for the regions of interest. The hatched band represents statistical and theoretical uncertainties on the backgrounds. The arrow indicates the VR selection on the variable shown. The bottom panel of each plot shows the ratio between the data and the post-fit MC prediction. The last bin includes the overflow events.
}
\label{fig:VR_distributions}
\end{figure}

The observed data yield and the post-fit background prediction for each CR and VR are shown in \figref{regionYields}. %
Good agreement is seen in all validation regions, with differences between the data and SM prediction within $1\sigma$.
The modeling of the extrapolated variable for each VR is shown in \figref{VR_distributions}, demonstrating good agreement in the shape
of the variables of interest. The resulting normalization factors for the \ttbar, single top, and \Zjet backgrounds are  $0.88\pm0.03$ for \ttbar,  $1.35\pm0.24$ for single top, and  $0.88\pm0.08$ for \Zjets. Normalization factors are obtained from the background-only fit (constrained only by CRs) considering all systematic uncertainties
for each of the three main backgrounds.

% End of text imported from the .//texinclude/bkg_estimation.tex input file

\FloatBarrier
\section{Systematic uncertainties}
\label{sec:systs}
%

% The next lines are included from the .//texinclude/systs.tex input file
\label{sec:systs}
Systematic uncertainties in the signal and background predictions arise
from theoretical uncertainties in the expected yield and MC modeling, and from experimental sources.
Experimental uncertainties reflect the precision of the energy and momentum calibration of jets and leptons, as well as corrections for the identification and reconstruction efficiencies in MC simulation.
The dominant experimental uncertainties on the signal efficiency are related to jets, including those in the jet energy scale and resolution~\cite{JETM-2018-05}
and the calibration of the $b$-tagging efficiency~\cite{FTAG-2018-01,FTAG-2020-08,FTAG-2019-02}.
The largest experimental uncertainties in the fitted background prediction in the SR are from the
mistagging of light-flavor jets and the jet energy resolution.
The experimental uncertainties associated with leptons each have a small impact on the final measurement, and include uncertainties in
the efficiency, energy scale and resolution of electrons~\cite{EGAM-2018-01} and muons~\cite{MUON-2018-03,MUON-2022-01},
and the calibration of the lepton trigger, identification, reconstruction, and isolation efficiencies.

Theoretical and MC modeling uncertainties of the \ttbar and single top backgrounds account for the choice of event generator, underlying-event tune, and their parameters.
The uncertainties are derived separately for each background process and are treated as uncorrelated nuisance parameters.
For \ttbar and single top backgrounds the theory systematics uncertainties considered arise from renormalization and factorization scale choices, PDF, $\alpha_s$, MC generator, and parton shower modeling.
For the \Zjets background only uncertainties from renormalization and factorization scale choices, PDF and $\alpha_s$ are considered.
As the normalization for the \ttbar, single top and \Zjets background is constrained in the likelihood fits, the uncertainties are derived from the normalization factor of the relevant CR
to each SR bin independently by comparing CR-to-SR yield ratios in alternative models.
The uncertainty in the background estimate due to the choice of MC event generator is estimated for \ttbar and single top backgrounds
by comparing the CR-to-SR yield ratios derived using the nominal \ttbar sample with another event sample produced with the
\MGNLO[2.6.2] generator at NLO in QCD using the five-flavor
scheme and the \NNPDF[2.3nlo] PDF set. The events were
interfaced with \PYTHIA[8.230], using the A14
set of tuned parameters and the \NNPDF[2.3lo]
PDF.
The uncertainty due to the parton shower and hadronization model was
evaluated by comparing the nominal sample of events with a sample where
events generated with the
\POWHEGBOX[v2]
generator were interfaced to
\HERWIG[7.04]~\cite{Bahr:2008pv,Bellm:2015jjp}, using the H7UE set
of tuned parameters~\cite{Bellm:2015jjp} and the \MMHT[lo] PDF set~\cite{Harland-Lang:2014zoa}.

An uncertainty in the single top yield due to the destructive interference between the \ttbar and \Wt processes is estimated by
using inclusive $WWbb$ samples generated at LO using \MGNLO split by different requirements on the presence of resonant top quarks.
The yields for each set of requirements is calculated in each SR bin and compared to one another to calculate a relative uncertainty on the \Wt yields.
This relative uncertainty is applied as an uncertainty on the transfer of the normalization factor from the CRst to each SR bin.
For background processes that are not normalized in the CRs (\Wjets, \ttV, diboson, triboson), the theoretical uncertainty on the NLO cross-section is applied.

Uncertainties in the expected yields of the stop signal samples are derived for each $m_{b\ell}^{0}$ bin of the SR as the envelope of variations to the renormalization and factorization scales, $\alpha_S$, and PDF set. A theoretical uncertainty on the stop cross-section is independently applied by varying the signal yield in all regions, and ranges from 8\% to 22\% for stop mass values from 600~\GeV to 1900~\GeV.
Uncertainties that affect the signal acceptance include the electron efficiency, with uncertainties calculated between 3\% and 4\% for the various \Stop masses when assuming $\BRe=\BRm=50\%$, and are between 5\% and 8\% when assuming $\BRe=100\%$.
Similarly, the muon efficiency uncertainties are between 2\% and 4\% when assuming $\BRe=\BRm=50\%$, and rise to 6\% when assuming $\BRm=100\%$.
The electron, muon, and jet energy scale and resolution uncertainties are generally below $1\%$ for the \Stop signal models,
reaching 1\% for masses near the \mbl threshold of 800~\GeV.
The $b$-tagging efficiency uncertainties are between 1\% and 3\%, reaching the largest value for the 600~\GeV signal model.

% End of text imported from the .//texinclude/systs.tex input file

\FloatBarrier
\section{Results}
\label{sec:result}
%

% The next lines are included from the .//texinclude/results.tex input file
\label{sec:results}
To evaluate the presence of signal in this analysis, a multi-bin fit is used to maximize the likelihood function $\mathcal{L}(\mu_\textrm{sig},\theta)$. The likelihood function is a product of Poisson distributions for each bin and depends on the parameter of interest, $\mu_\textrm{sig}$. This parameter, referred to as the signal strength, is defined by the ratio of the fitted signal cross-section versus the theoretical cross-section. The effect of systematic uncertainties in the signal and background predictions are described by nuisance parameters subject to Gaussian penalty terms in the likelihood function.
For model dependent results, two exclusion fits were performed for each lepton branching ratio point probed: one inclusive 15-bin fit, which is agnostic to lepton final state, and one 45-bin fit simultaneously across all three lepton flavor channels $(e\mu ,\mu\mu ,ee)$ as separate SRs.
For each stop mass and branching ratio, the configuration that provides the strongest expected limit on the signal strength will be used.
The flavor-agnostic configuration typically has the stronger expected sensitivity when \BRe, \BRm, and \BRt are similar in size,
while the flavor-aware configuration is more sensitive in the corners of the branching ratio plane.

\begin{figure}[h]
\centering
\subfigure[]{%
\includegraphics[width=0.47\textwidth]{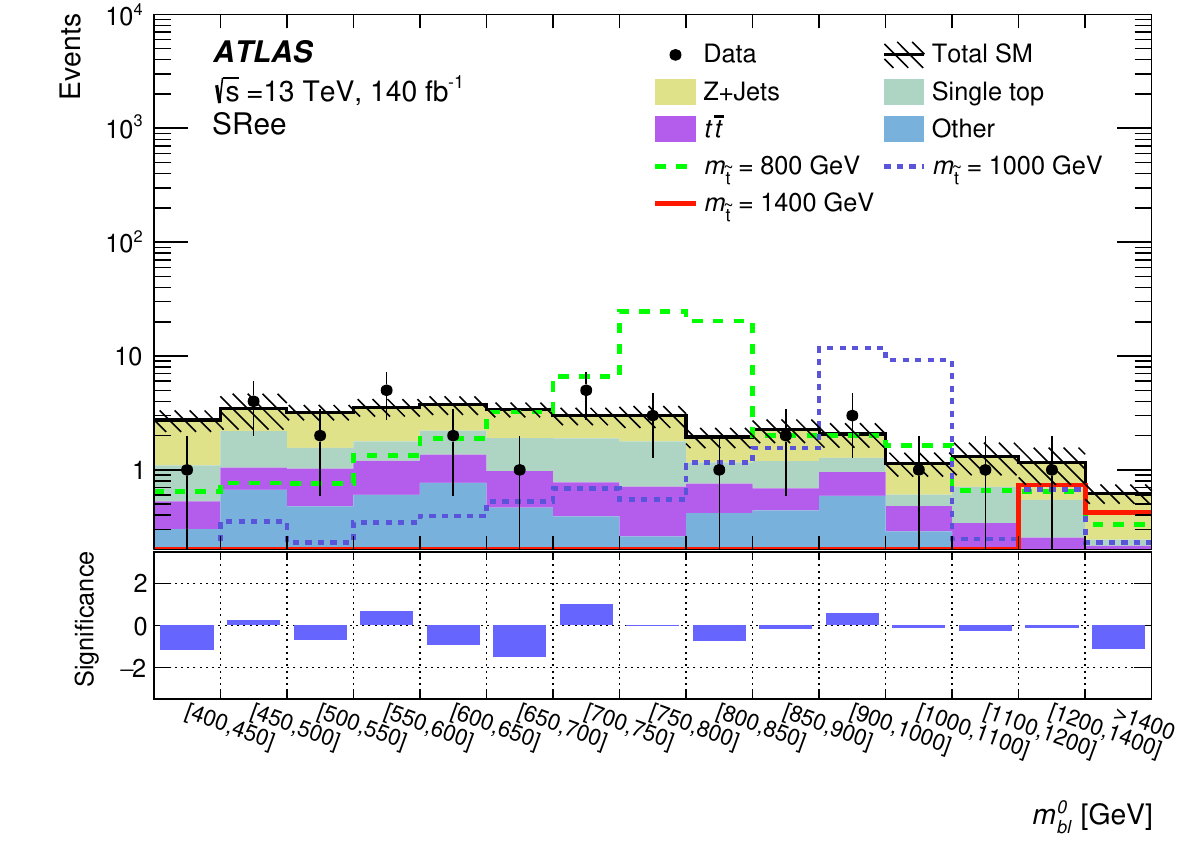}
\label{fig:SRPulls45Bin:ee}
}
\subfigure[]{%
\includegraphics[width=0.47\textwidth]{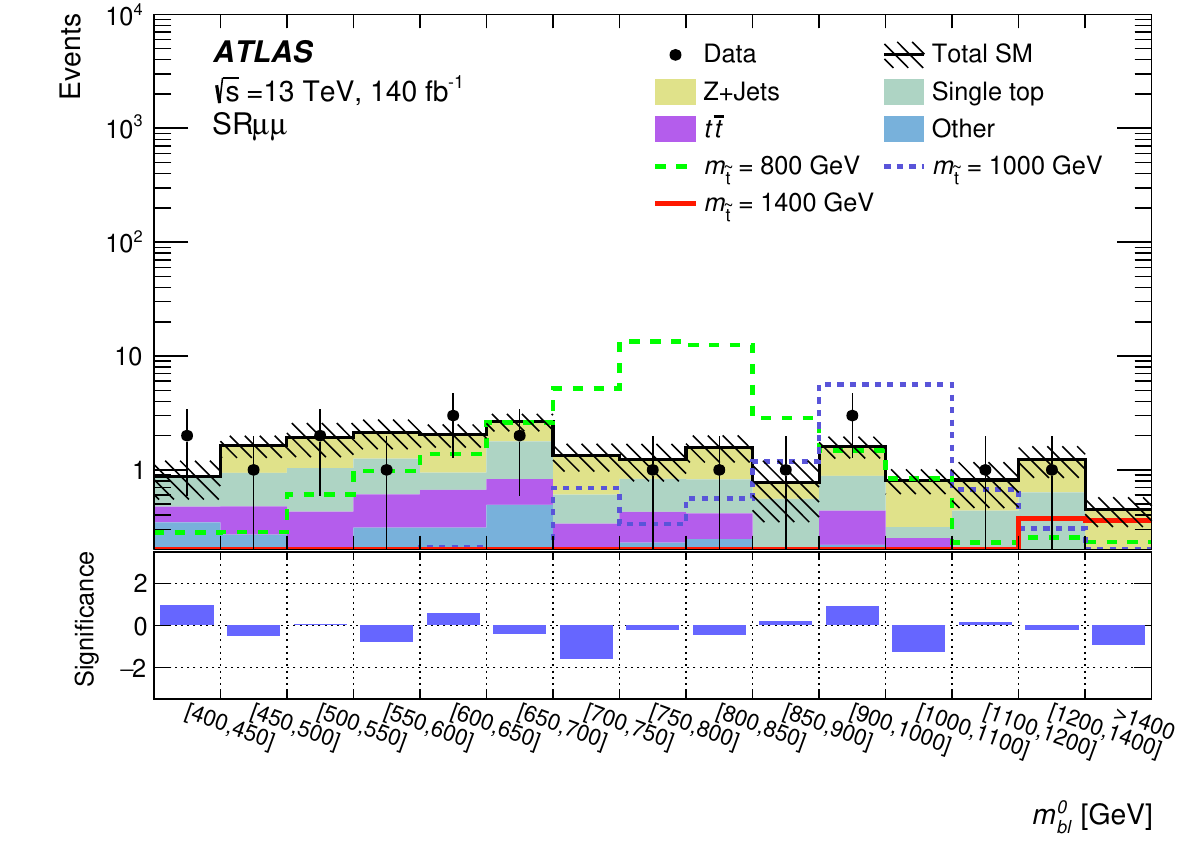}
\label{fig:SRPulls45Bin:mm}
}
\subfigure[]{%
\includegraphics[width=0.47\textwidth]{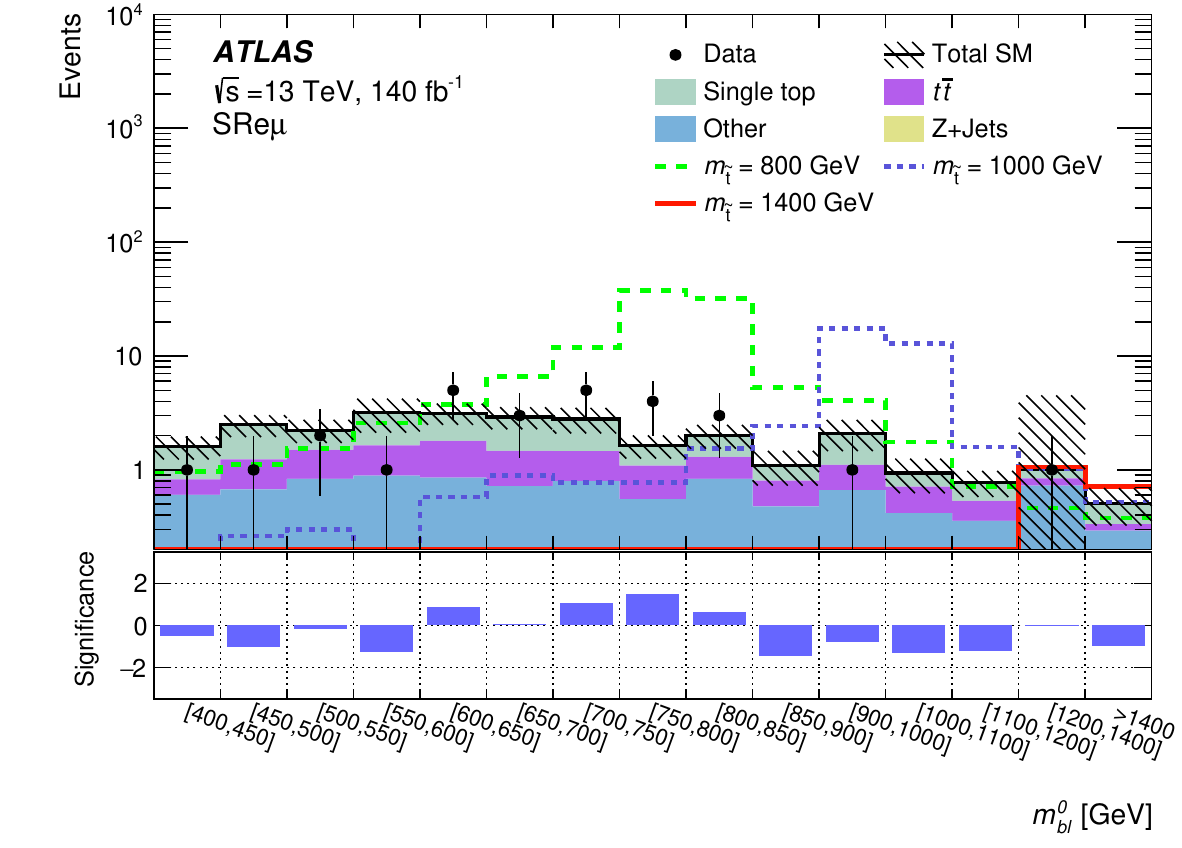}
\label{fig:SRPulls45Bin:em}
}
\caption{The observed data and post-fit Monte Carlo expectation for SM processes in each SR bin for the flavor-aware configuration in \subref{fig:SRPulls45Bin:ee} the $ee$ channel, \subref{fig:SRPulls45Bin:mm} the $\mu\mu$ channel, and \subref{fig:SRPulls45Bin:em} the $e\mu$ channel. Each plot shows the yields for a single flavor channel of the 45-bin background-only fit. All systematic uncertainties are applied in this fit. The "Other" category consists of \Wjets, \ttV, diboson and triboson processes. Processes are listed from largest to smallest contribution to the given region reading left to right and then top to bottom. The bottom panel of each plot shows the significance of the difference between the expected yields and observed data. Expected signal contributions for different stop mass hypotheses are shown, assuming equal branching ratios to electron, muon and $\tau$-leptons. The significance is computed through a profile likelihood method~\cite{Cousins_2008}.}
\label{fig:SRPulls45Bin}
\end{figure}

\begin{figure}[h]
\centering
\includegraphics[width=0.9\textwidth]{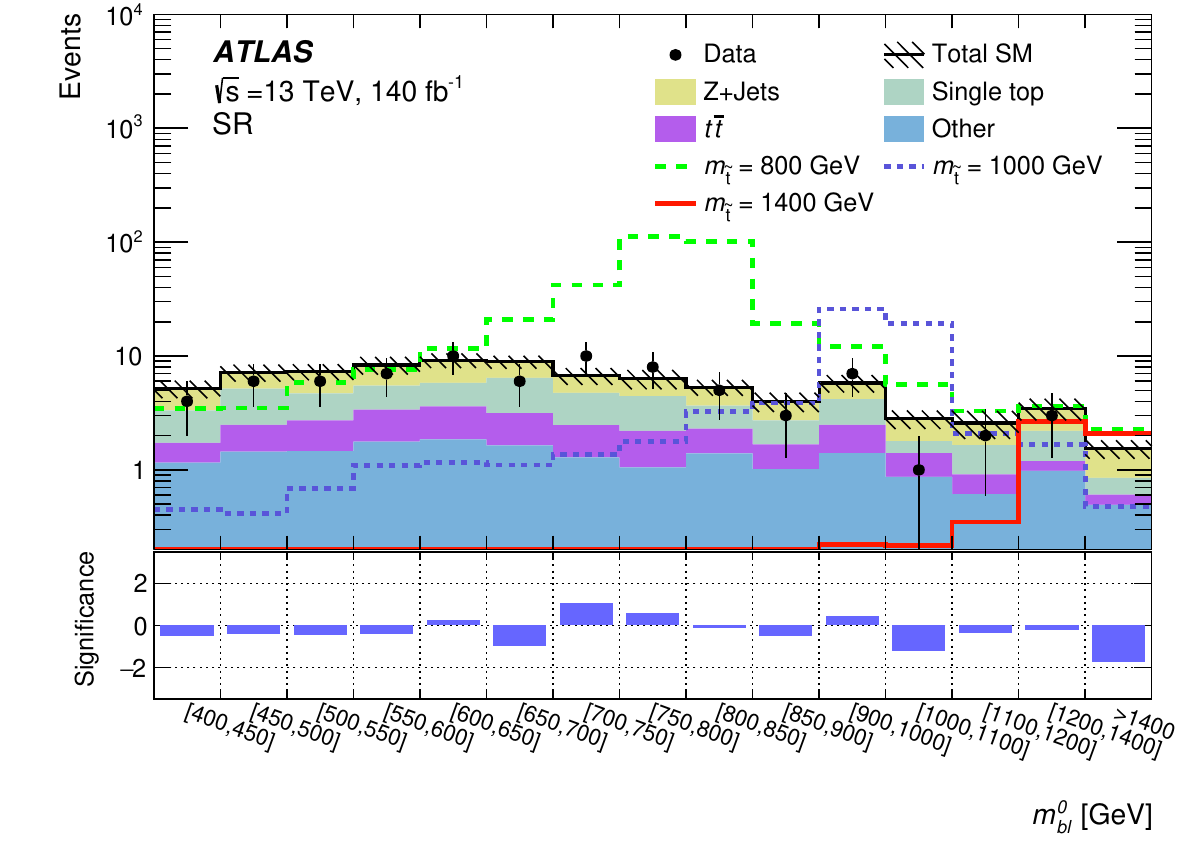}
\caption{The observed data and post-fit Monte Carlo expectation for SM processes in each SR bin for the flavor-agnostic configuration. The results are obtained from the SRs using the 15-bin background-only fit. All systematic uncertainties are applied in this fit. The "Other" category consists of \Wjets, \ttV, diboson and triboson processes. Processes are listed from largest to smallest contribution to the given region reading left to right and then top to bottom. The bottom panel of the plot shows the significance of the difference between the expected yields and observed data. Expected signal contributions for different stop mass hypotheses are shown assuming equal branching ratios to electron, muon and $\tau$-leptons. The significance is computed through a profile likelihood method~\cite{Cousins_2008}.}
\label{fig:SRPulls15Bin}
\end{figure}

\begin{figure}[h]
\centering
\subfigure[]{%
\includegraphics[width=0.31\textwidth]{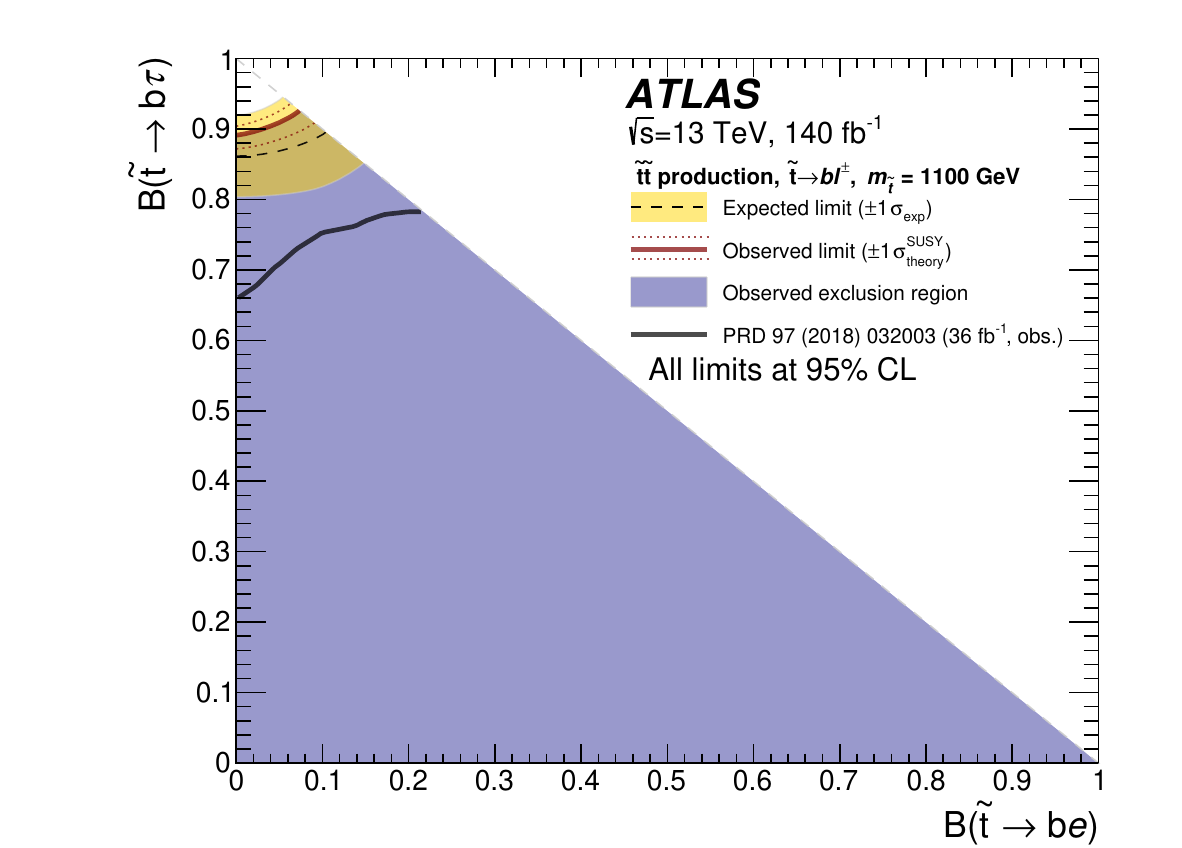}
}
\subfigure[]{%
\includegraphics[width=0.31\textwidth]{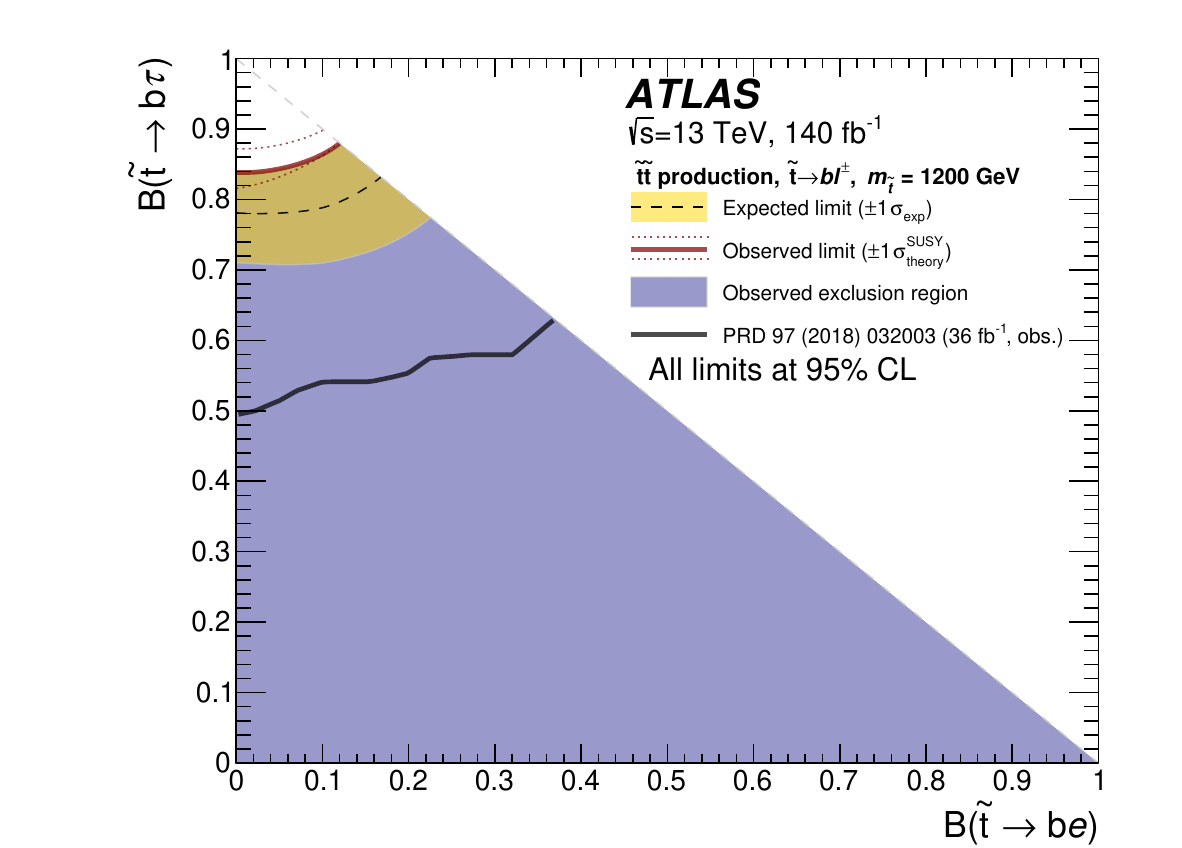}
}
\subfigure[]{%
\includegraphics[width=0.31\textwidth]{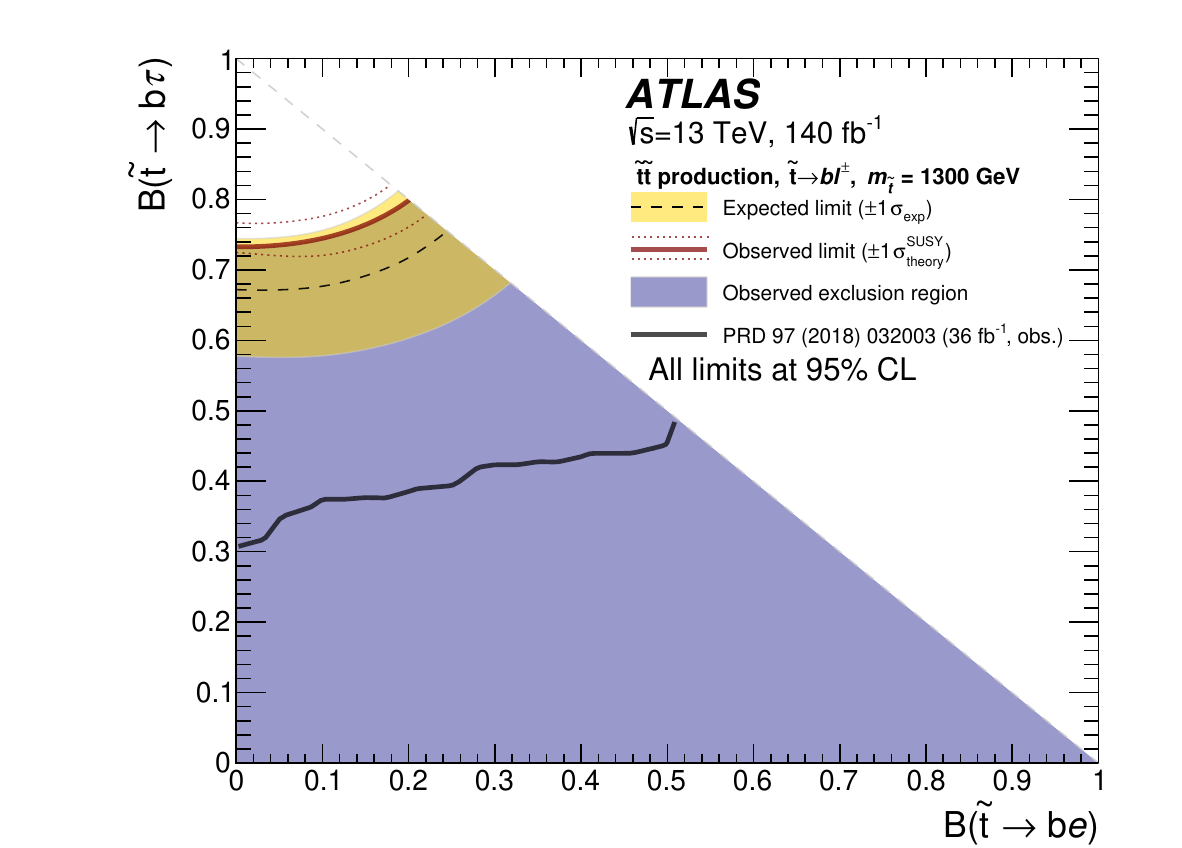}
}
\subfigure[]{%
\includegraphics[width=0.31\textwidth]{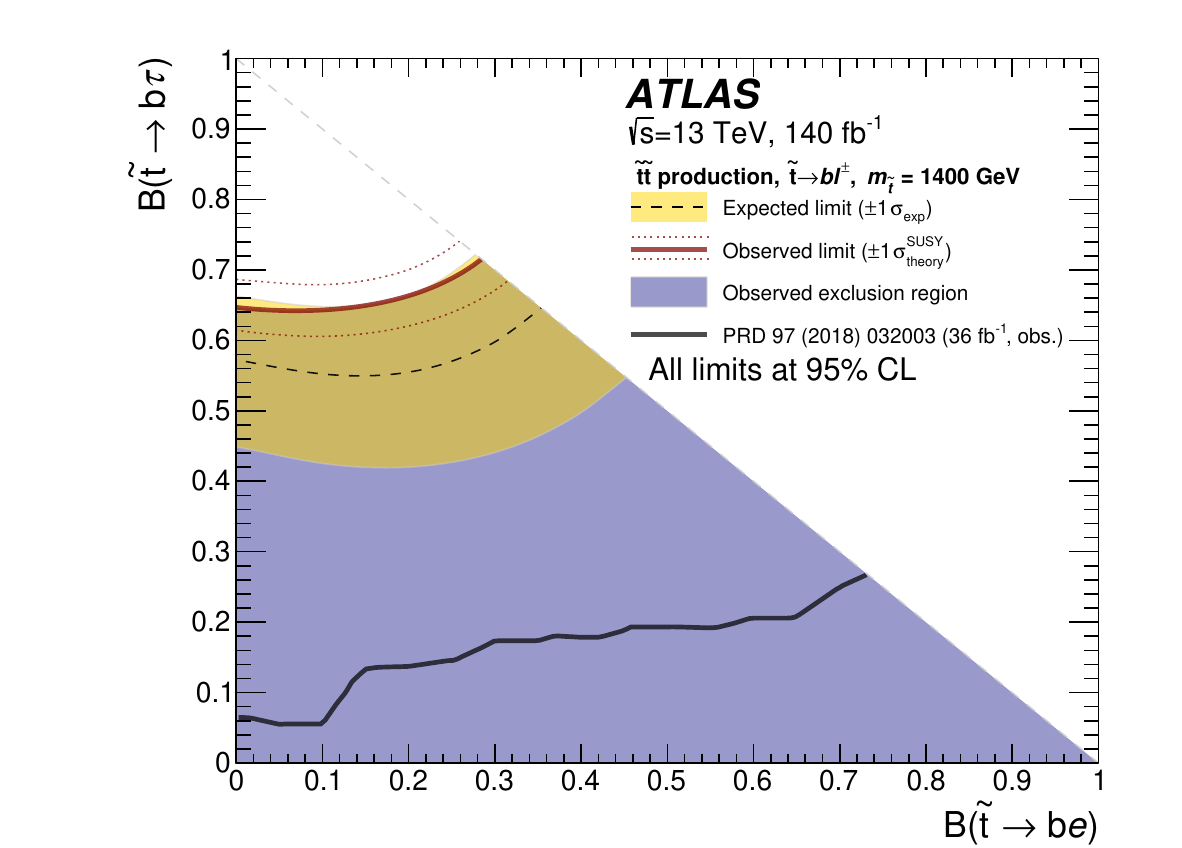}
}
\subfigure[]{%
\includegraphics[width=0.31\textwidth]{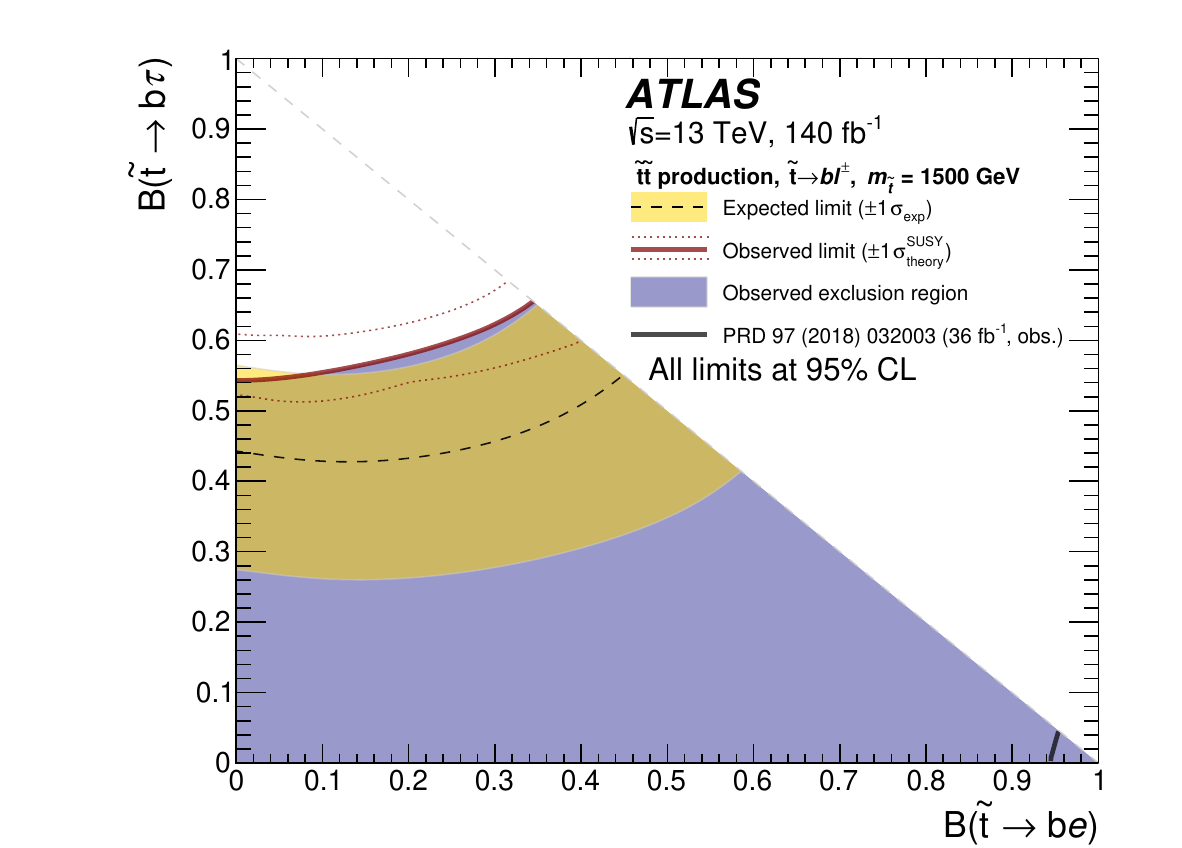}
}
\subfigure[]{%
\includegraphics[width=0.31\textwidth]{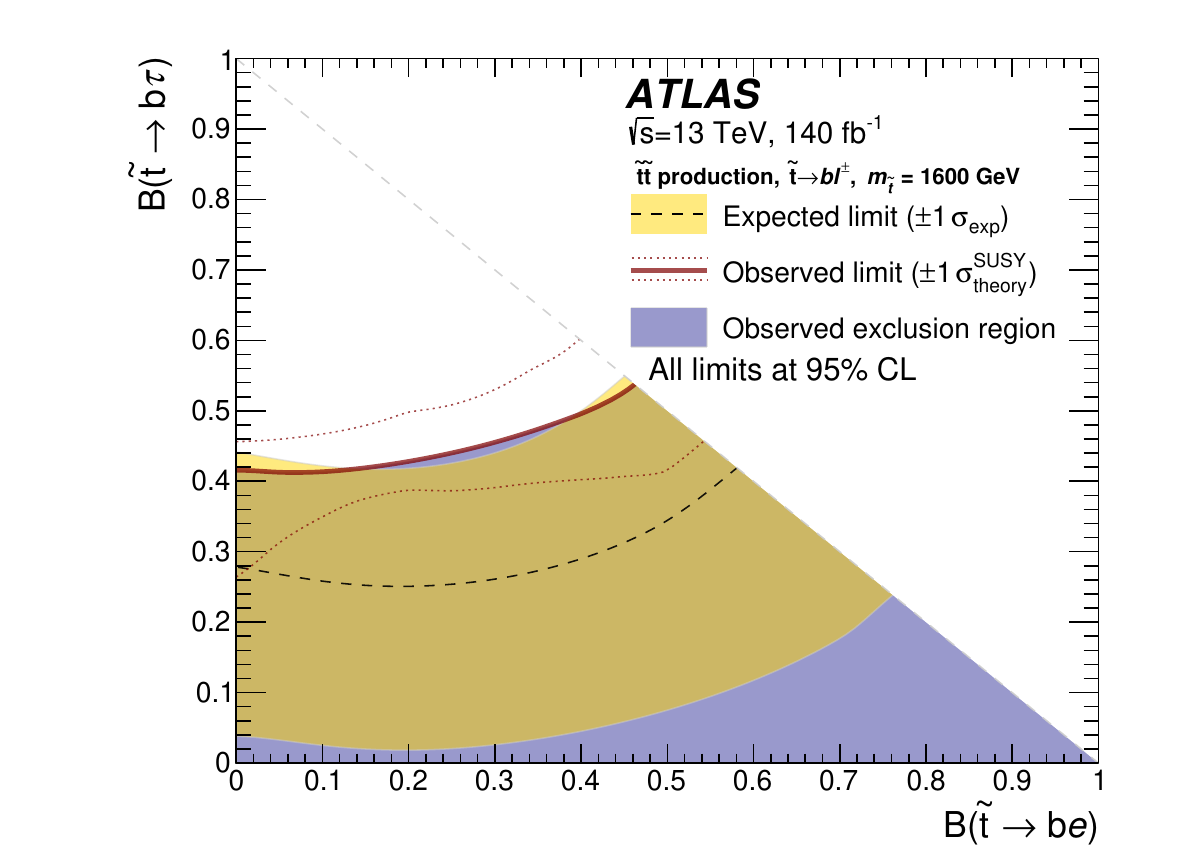}
}
\subfigure[]{%
\includegraphics[width=0.31\textwidth]{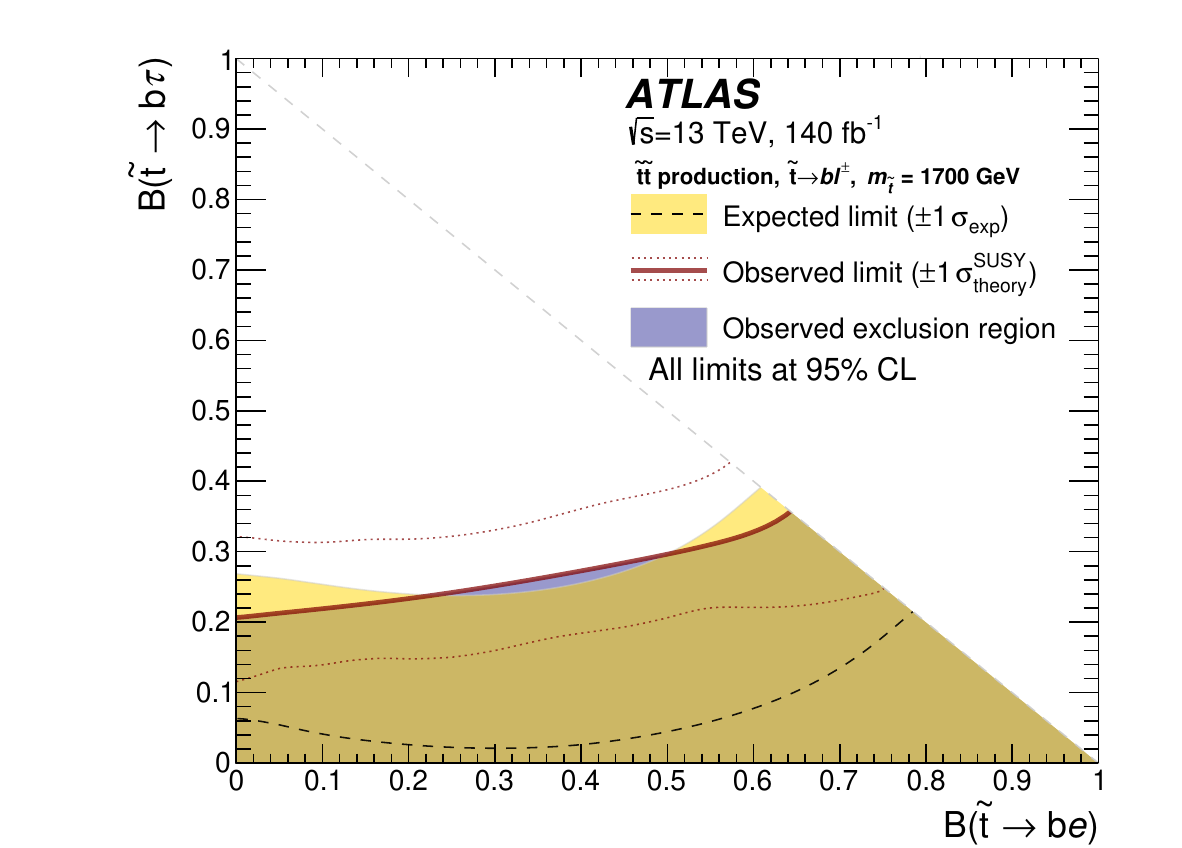}
}
\subfigure[]{%
\includegraphics[width=0.31\textwidth]{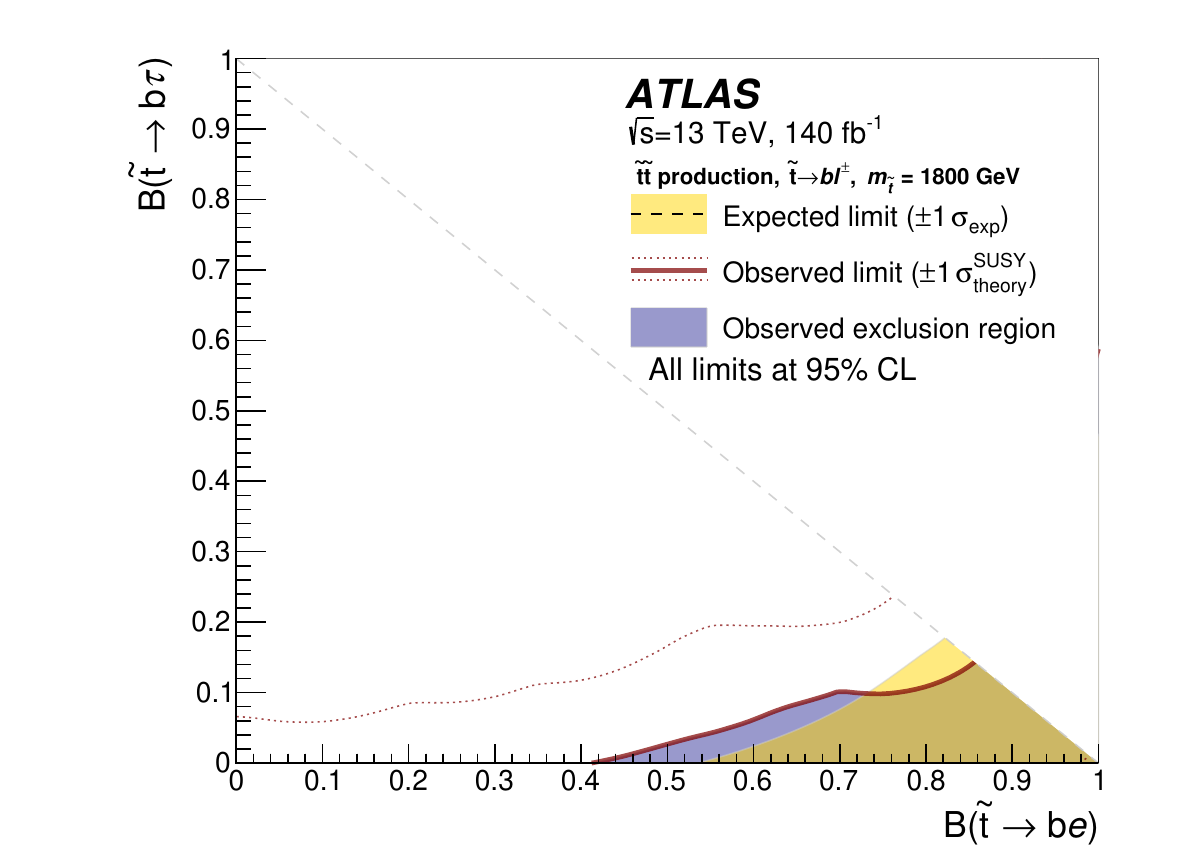}
}
\subfigure[]{%
\includegraphics[width=0.31\textwidth]{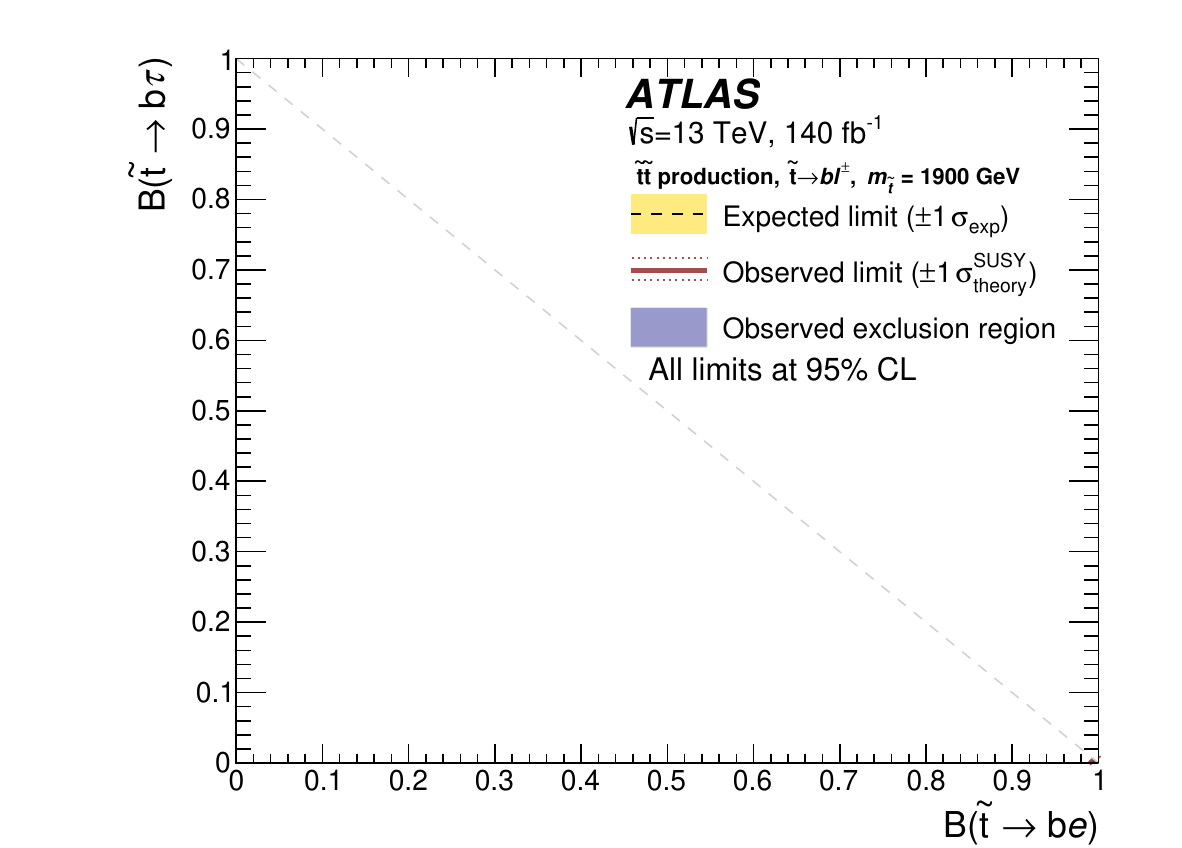}
}
\caption{Observed (solid red) and expected (dashed black) 95\% CL exclusions in the \BRt vs \BRe plane
for various mass values between 1100 and 1900~\GeV (a)-(i).
The sum of \BRe, \BRm, and \BRt is assumed to be unity everywhere; hence, the \BRm can be inferred for each point in the figure.
The yellow band reflects the $\pm 1\sigma$ uncertainty of the expected limit due to theoretical, experimental, and MC statistical uncertainties.
The shaded purple area represents the branching ratios that are observed to be excluded.
The dotted red lines correspond to the $\pm 1\sigma$ cross-section uncertainty of the observed limit
derived by varying the signal cross-section by the theoretical uncertainties. The black solid line represents previous limits placed by the early Run 2 analysis~\cite{SUSY-2016-29}. In figures that do not have previous limits shown, there was no sensitivity for this mass range in the previous analysis. The chosen branching ratio plane to interpret the results allows an easier comparison with the results of the scan over the neutrino mass hierarchy presented in Ref.~\cite{Marshall:2014kea}. }
\label{fig:limit_contours}
\end{figure}

The observed yields and fitted background predictions for the SRs are shown in \figref{SRPulls45Bin} for the 45-bin fit, and in \figref{SRPulls15Bin} for the 15-bin fit.
Yields for the 45-bin SR configurations are split across three plots (one for each lepton flavor channel).
Estimated background is in agreement with observed data within $1\sigma$ over almost all signal bins. Validation regions show that background is modeled well.
Model-dependent limits are given in \figref{limit_contours} for a range of assumed stop masses.
The model-dependent results show observed limits that are stronger than the expected limits at the $1\sigma$ level, primarily due to the highest mass signal bin where zero events are observed while 2.14 $\pm 0.60$ events are expected from background predictions.
The behavior of the limits shown in \figref{limit_contours} is driven by the highest mass bin of the SR, which has low statistics but an observed deficit.

\begin{table}
\caption[Model-independent limits from Discovery fit]{
Left to right: the number of events expected by the SM, the number of events observed, 95\% CL upper limits on the visible cross-section
($\langle\epsilon\sigma\rangle_{\textrm obs}^{95}$) and on the number of
signal events ($S_{\textrm obs}^{95}$ ).  The fifth column
($S_{\textrm exp}^{95}$) shows the 95\% CL upper limit on the number of
signal events, given the expected number (and $\pm 1\sigma$
excursions on the expectation) of background events.
The last column
indicates the $CL_b$ value, i.e. the confidence level observed for
the background-only hypothesis. The discovery $p$-value ($p(s = 0)$) is 0.50 with significance (Z) of 0 for all presented mass ranges. The discovery limits were derived using 10,000 pseudo-experiments. }
\centering
\setlength{\tabcolsep}{0.0pc}
\begin{tabular*}{\textwidth}{@{\extracolsep{\fill}}lccccccc}
\noalign{\smallskip}\hline\noalign{\smallskip}{Mass Range [GeV]}                       & $N^{\text{SM}}_{\text{exp}}$ & $N_{\text{obs}}$  & $\langle\epsilon{\sigma}\rangle_{\text{obs}}^{95}$[fb]  &  $S_{\text{obs}}^{95}$  & $S_{\text{exp}}^{95}$ & $CL_{b}$\\
\noalign{\smallskip}\hline\noalign{\smallskip}
$\ge 400$    & $84.0 \pm 8.2$  & $78$  & $0.16$ &  $22$ & $ { 32 }^{ +11 }_{ -8 }$ & $0.09$\\%
$\ge 450$    & $81.8 \pm 7.6$  & $74$  & $0.13$ &  $19$ & $ { 29 }^{ +10 }_{ -8 }$ & $0.06$\\%
$\ge 500$    & $72.7 \pm 7.9$  & $68$  & $0.15$ &  $22$ & $ { 30 }^{ +10 }_{ -7 }$ & $0.12$\\%
$\ge 550$    & $67.7 \pm 7.3$  & $62$  & $0.13$ &  $18$ & $ { 27 }^{ +9 }_{ -7 }$ & $0.11$\\%
$\ge 600$    & $59.9 \pm 6.8$  & $55$  & $0.13$ &  $18$ & $ { 25 }^{ +9 }_{ -6 }$ & $0.12$\\%
$\ge 650$    & $50.2 \pm 5.9$  & $45$  & $0.11$ &  $15$ & $ { 22 }^{ +8 }_{ -6 }$ & $0.11$\\%
$\ge 700$    & $43.4 \pm 5.4$  & $39$  & $0.10$ &  $14$ & $ { 20 }^{ +7 }_{ -5 }$ & $0.15$\\%
$\ge 750$    & $33.3 \pm 4.6$  & $29$  & $0.08$ &  $12$ & $ { 17 }^{ +6 }_{ -5 }$ & $0.12$\\%
$\ge 800$    & $26.6 \pm 3.6$  & $21$  & $0.06$ &  $8.9$ & $ { 14 }^{ +6 }_{ -4 }$ & $0.07$\\%
$\ge 850$    & $20.1 \pm 3.2$  & $16$   & $0.06$ &  $8.3$ & $ { 13 }^{ +5 }_{ -3 }$ & $0.10$\\%
$\ge 900$    & $16.9 \pm 2.8$  & $13$   & $0.05$ &  $7.4$ & $ { 11 }^{ +5 }_{ -3 }$ & $0.11$\\%
$\ge 1000$    & $9.9 \pm 1.8$  & $6$     & $0.04$ &  $5.1$ & $ { 8.4 }^{ +4 }_{ -3 }$ & $0.08$\\%
$\ge 1100$    & $8.5 \pm 1.6$  & $5$     & $0.03$ &  $4.9$ & $ { 7.8 }^{ +4 }_{ -2 }$ & $0.09$\\%
$\ge 1200$    & $5.3 \pm 1.3$  & $3$     & $0.03$ &  $4.2$ & $ { 6.4 }^{ +3 }_{ -2 }$ & $0.14$\\%
$\ge 1400$    & $1.8 \pm 0.5$  & $0$   & $0.02$ &  $3.0$ & $ { 4.2 }^{ +2 }_{ -1 }$ & $0.11$ &\\%
\noalign{\smallskip}\hline\noalign{\smallskip}
\end{tabular*}
\label{tab:UnblindedDiscoveryResults}
\end{table}

\begin{figure}[!h]
\centering
\includegraphics[width=0.6\textwidth]{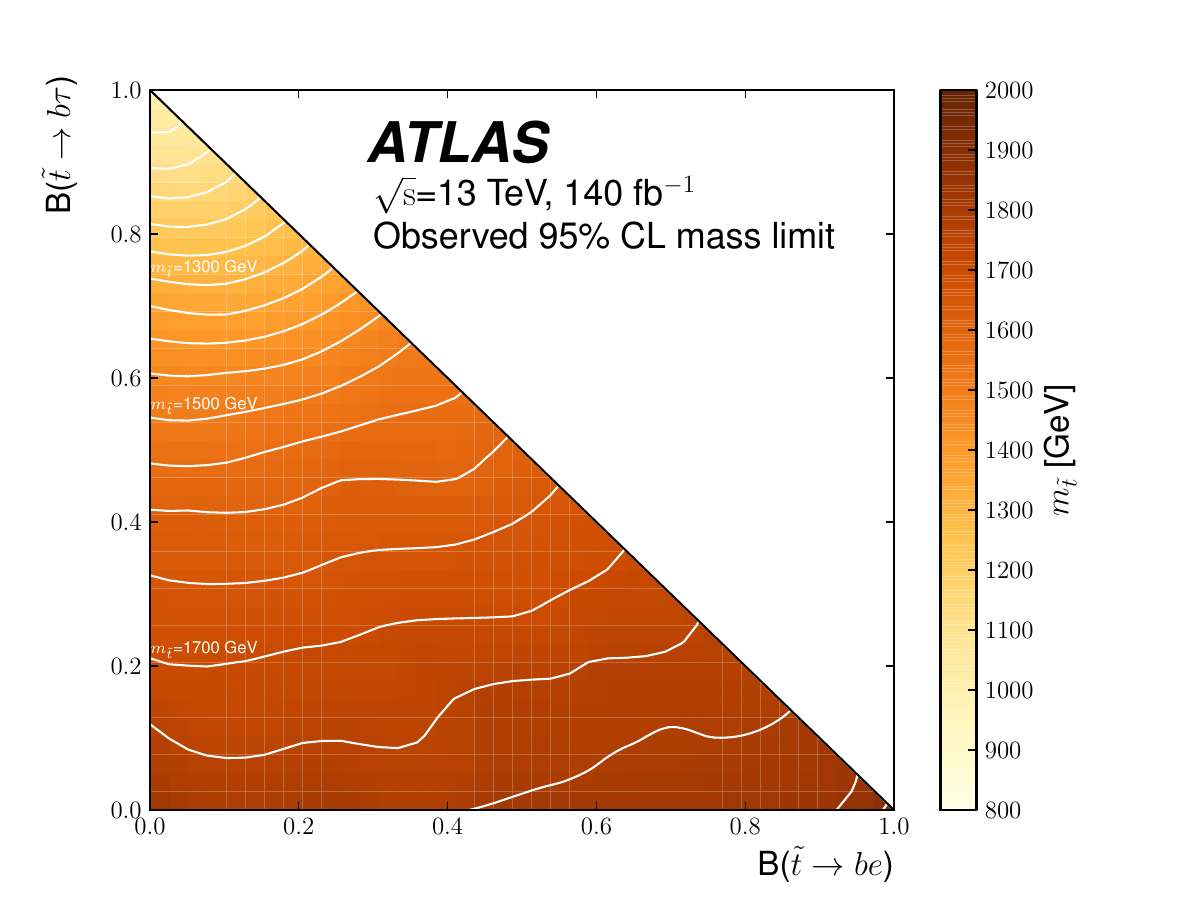}
\caption{The observed lower limits on the \StopSym mass at 95\% CL as a function of \StopSym branching ratios.
The sum of \BRe, \BRm, and \BRt is assumed to be unity everywhere; hence, the \BRm can be inferred for each point in the figure.
As the branching ratio \BRt increases, the
observed number of events with electrons or muons in the final state
decreases, reducing the sensitivity of the search. Benchmark stop masses are provided in 50 GeV intervals.
}
\label{fig:mass_limit_obs}
\end{figure}

\begin{figure}[h]
\centering
\subfigure[]{%
\includegraphics[width=0.47\textwidth]{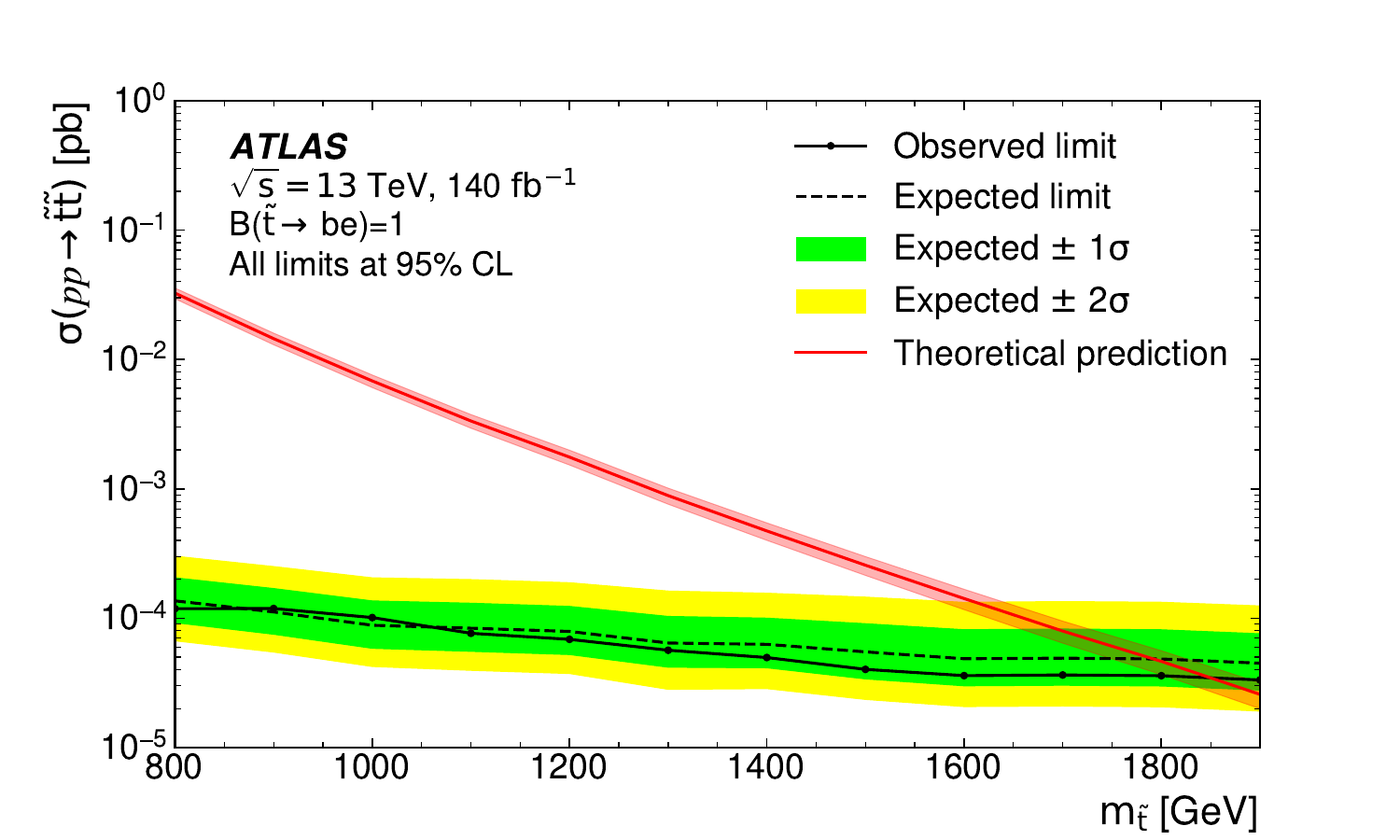}
\label{fig:cross:100e}
}
\subfigure[]{%
\includegraphics[width=0.47\textwidth]{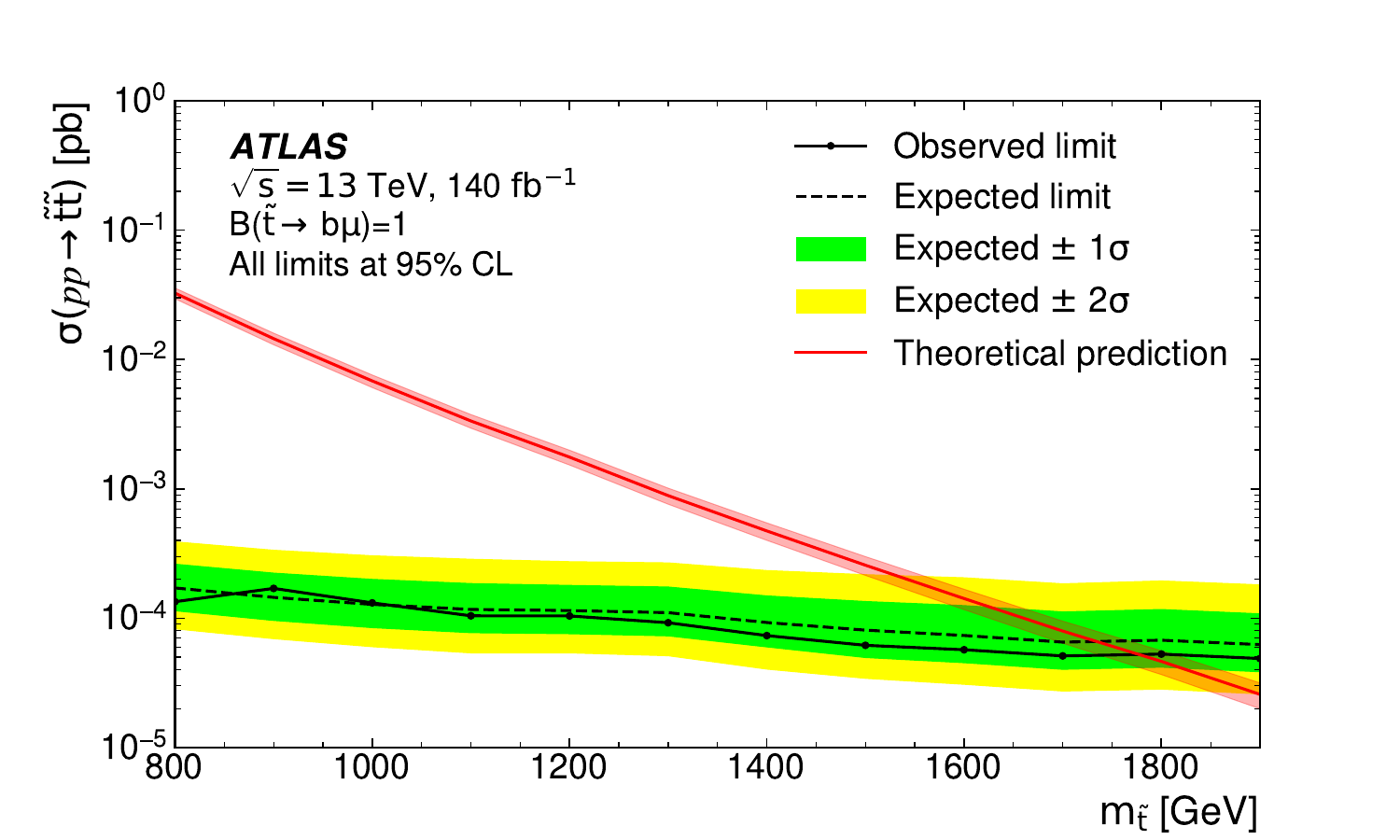}
\label{fig:cross:100m}
}
\subfigure[]{%
\includegraphics[width=0.47\textwidth]{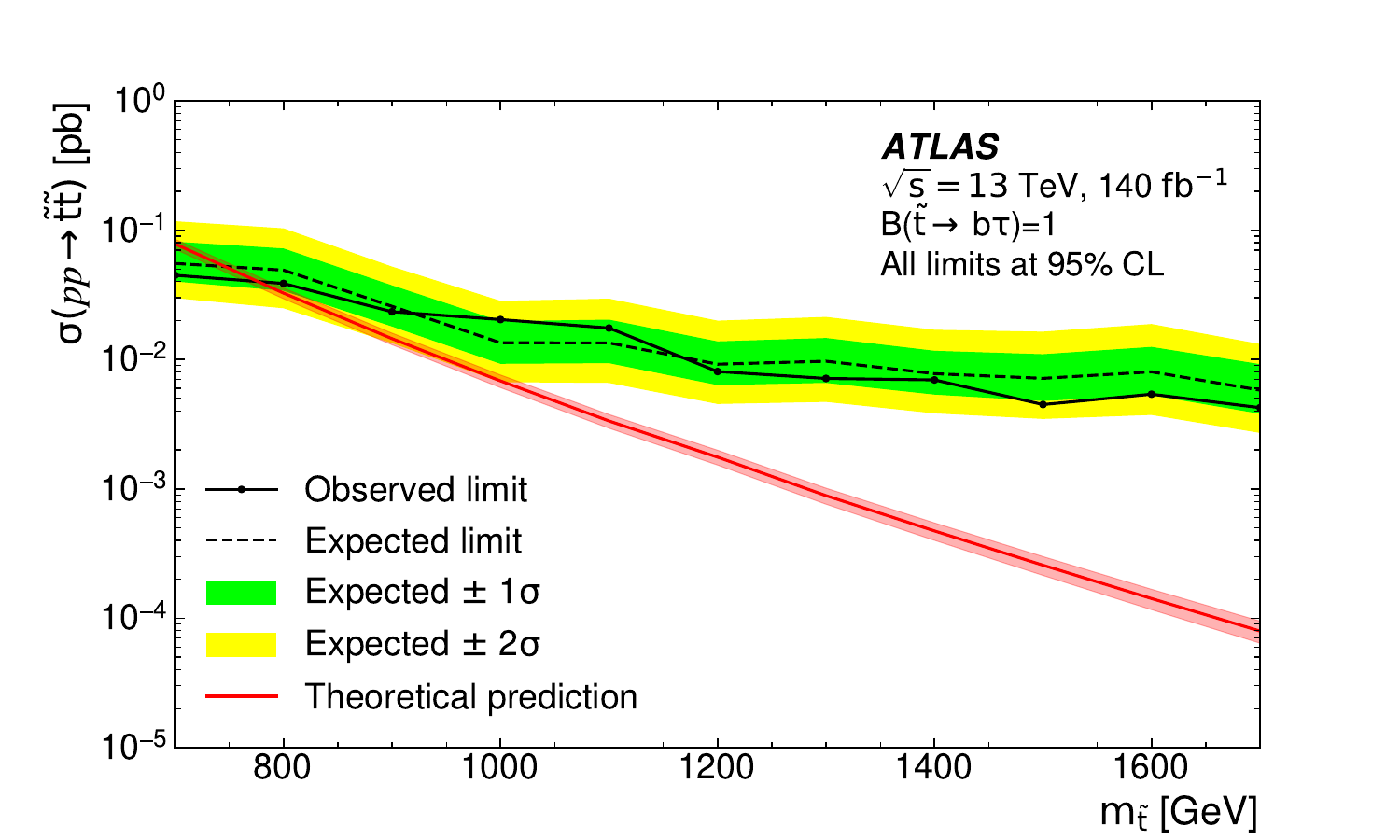}
\label{fig:cross:100t}
}
\subfigure[]{%
\includegraphics[width=0.47\textwidth]{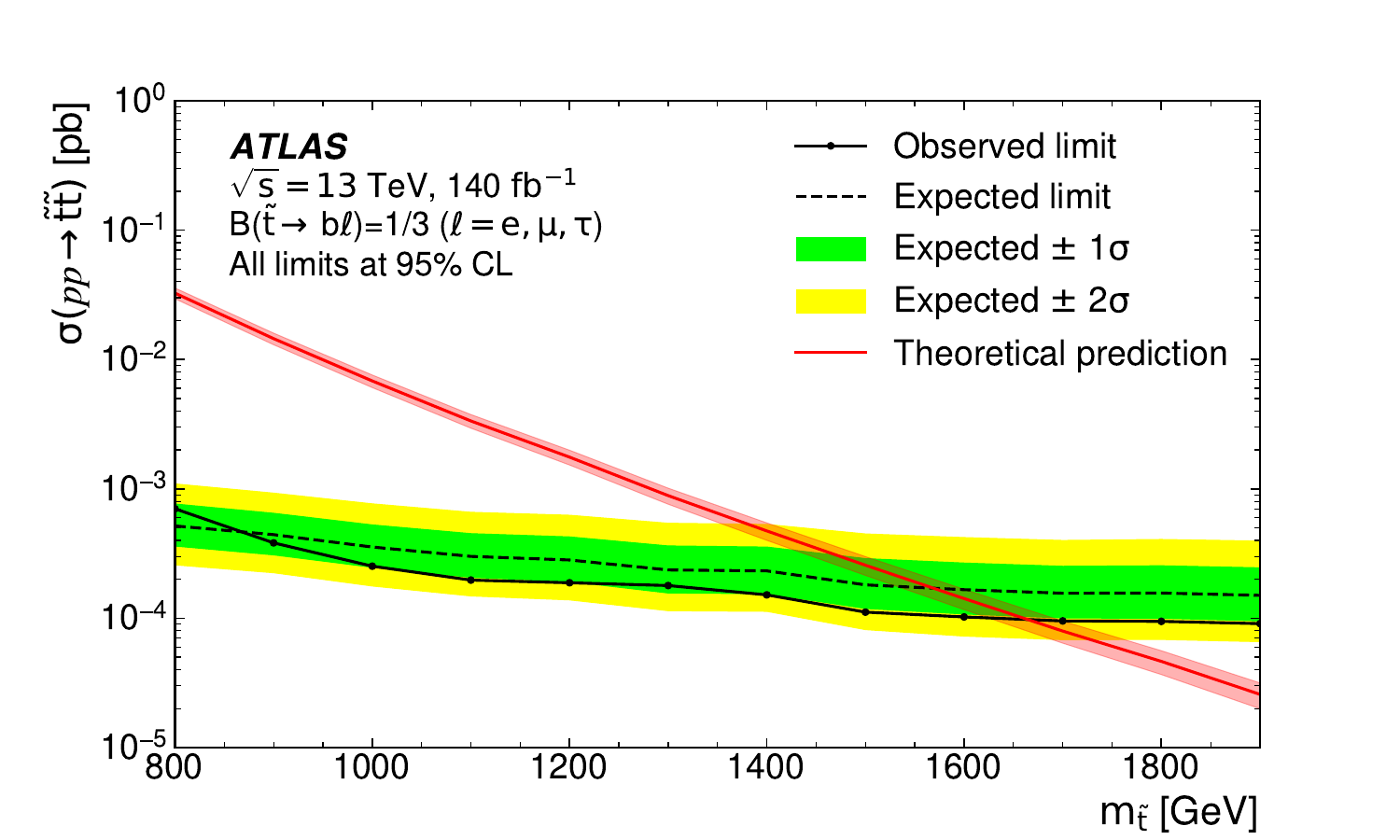}
\label{fig:cross:dem}
}
\caption{Observed (solid black) and expected (dashed black) 95\% CL upper limits on the stop pair production cross-section as a function of stop mass for different branching ratio hypotheses:\subref{fig:cross:100e} 100\% to a $b$-quark and an electron, \subref{fig:cross:100m} 100\% to a $b$-quark and a muon, \subref{fig:cross:100t} 100\% to a $b$-quark and a tau, and \subref{fig:cross:dem} democratic branching ratios for this decay. The red line represents theory prediction for the stop cross-section and corresponding uncertainty in a lighter red band~\cite{Beenakker_2016}.
}
\label{fig:cross}
\end{figure}

Exclusion limits are derived at 95\% CL for stop pair production.
Limits are obtained through an exclusion fit based on a profile log-likelihood ratio test using the \CLs prescription~\cite{Read:2002hq} and asymptotic approximation~\cite{Cowan:2010js},
following the simultaneous fit to the CRs and SRs~\cite{Baak:2014wma}.
The signal contributions in both the SRs and CRs are accounted for in the fit, although they are negligible in the latter.
Exclusion fits are performed separately for various branching ratio assumptions, sampling, in steps of 5\%,
values of \BRe, \BRm, and \BRt whose sum is unity and reweighting events in the signal samples according to the generated decays.
The limits are strongest at low values of \BRt, where the expected number of
events with electrons or muons in the final state is largest.
Expected limits are slightly stronger for increasing \BRe, reflecting a higher
trigger efficiency for electrons than for muons.
Stops with \BRt up to 100\% are excluded for masses below 800 ~\GeV, while
those with \BRm up to 100\% are excluded up to 1800~\GeV and those with \BRe up to 100\% are excluded up to 1900~\GeV.
Observed limits are stronger than expected,
reflecting the lower-than-expected event yield at high \Mbl.
Exclusion contours reflecting the highest stop mass excluded at a 95\% CL for a given
point in the branching ratio plane are shown in \figref{mass_limit_obs}, which emphasizes this search's ability to probe different branching ratio regions. These branching ratio regions can inform theoretical predictions for the normal versus inverted neutrino mass ordering~\cite{Marshall:2014kea}.
One-dimensional limit plots for different branching ratio benchmarks are shown in \figref{cross}.
Cross-section limits for the stop decaying into a $b$-jet and lepton with democratic branching ratio to each of the lepton flavors without systematic uncertainties would be 2 times stronger for lower stop masses and 1.2 times stronger for higher stop masses.

In addition to using the aforementioned SR definitions for model-dependent results, a second set of regions were defined to produce model-independent results. These model-independent regions use the same baseline SR selections as the regions with the orthogonal bins in $\Mbl^0$ replaced by 15 inclusive, nested bins in $\Mbl^0$. Each nested bin starts at the lower bin edge of the respective nominal SR definition.
For each inclusive SR, model-independent upper limits are derived on the visible cross-section of potential beyond-the-SM (BSM) processes at a 95\% CL.
The visible cross-section is defined to be $\sigma \cdot \epsilon \cdot A$, where $\sigma$ is the cross-section, $\epsilon$ is the efficiency, and $A$ is the detector acceptance.
A likelihood fit is performed to the number of observed events in all three CRs and the target SR where a generic BSM process
is assumed to contribute to the SR only.
The $p$-value of this statistical test is obtained using pseudo-experiments. No theoretical or systematic uncertainties are considered for the potential BSM signal except the luminosity uncertainty from 2015 to 2018.
The observed ($S^{95}_\mathrm{obs}$) and expected ($S^{95}_\mathrm{exp}$) limits on the number of BSM events
are derived at 95\% CL inclusively across flavor channel to avoid assumptions on the underlying model.
These results are shown in \tabref{UnblindedDiscoveryResults}.
Also shown are the observed limits on the visible cross-section $\sigma_\mathrm{vis}$, defined as $S^{95}_\mathrm{obs}$ normalized to the integrated luminosity,
and representing the product of the production cross-section, acceptance, and selection efficiency of a generic BSM signal.
Limits on $\sigma_\mathrm{vis}$ are set between 0.02 fb and 0.16 fb, depending on the \Mbl requirement.

\FloatBarrier

% End of text imported from the .//texinclude/results.tex input file

\FloatBarrier
\section{Conclusion}
\label{sec:conclusion}
%

% The next lines are included from the .//texinclude/conclusion.tex input file
\label{secref:conclusion}
This paper presents the full Run 2 ATLAS results on the search for stop pair production, with each stop decaying
via an RPV coupling to a $b$-quark and a lepton.
The search uses 140 \ifb of $\sqrt{s}=13~\TeV$
proton--proton collision data collected with the ATLAS detector at the LHC from 2015 to 2018.
The resulting final state is characterized by two jets, at least one of which is $b$-tagged, and two light, opposite-charge leptons (electron or muon).
This signal is searched for in the distribution of the leading invariant mass of the lepton-jet resonance reconstructing the stop pair.
No significant excess of events over the SM prediction is observed,
and limits are set on the stop mass under different hypotheses for the branching ratios at 95\% confidence level.
Limits are set on stop masses up to 800~\GeV for a $b\tau$ branching ratio of 100\%, 1800~\GeV for a $b\mu$ branching ratio of 100\% and 1900~\GeV for a $be$ branching ratio of 100\%.
These results significantly extend the mass exclusion limits from previous searches.
The improvement in sensitivity is due to the larger dataset, performing a fit to the distribution of the mass of the leading lepton-jet pair, increasing the threshold for the highest bin from 1100 to 1400~\GeV, and reducing the impact of systematic uncertainties with improved control regions.
Model-independent upper limits are set on the cross-section of potential BSM processes inclusive in the $ee$, $e\mu$, and $\mu\mu$ channels and between 0.02 fb and 0.16 fb.

% End of text imported from the .//texinclude/conclusion.tex input file

\FloatBarrier
\section*{Acknowledgements}
%

% The next lines are included from the .//acknowledgements/Acknowledgements.tex input file
%
%

%
%

We thank CERN for the very successful operation of the LHC and its injectors, as well as the support staff at
CERN and at our institutions worldwide without whom ATLAS could not be operated efficiently.

The crucial computing support from all WLCG partners is acknowledged gratefully, in particular from CERN, the ATLAS Tier-1 facilities at TRIUMF/SFU (Canada), NDGF (Denmark, Norway, Sweden), CC-IN2P3 (France), KIT/GridKA (Germany), INFN-CNAF (Italy), NL-T1 (Netherlands), PIC (Spain), RAL (UK) and BNL (USA), the Tier-2 facilities worldwide and large non-WLCG resource providers. Major contributors of computing resources are listed in Ref.~\cite{ATL-SOFT-PUB-2023-001}.

We gratefully acknowledge the support of ANPCyT, Argentina; YerPhI, Armenia; ARC, Australia; BMWFW and FWF, Austria; ANAS, Azerbaijan; CNPq and FAPESP, Brazil; NSERC, NRC and CFI, Canada; CERN; ANID, Chile; CAS, MOST and NSFC, China; Minciencias, Colombia; MEYS CR, Czech Republic; DNRF and DNSRC, Denmark; IN2P3-CNRS and CEA-DRF/IRFU, France; SRNSFG, Georgia; BMBF, HGF and MPG, Germany; GSRI, Greece; RGC and Hong Kong SAR, China; ISF and Benoziyo Center, Israel; INFN, Italy; MEXT and JSPS, Japan; CNRST, Morocco; NWO, Netherlands; RCN, Norway; MNiSW, Poland; FCT, Portugal; MNE/IFA, Romania; MESTD, Serbia; MSSR, Slovakia; ARIS and MVZI, Slovenia; DSI/NRF, South Africa; MICIU/AEI, Spain; SRC and Wallenberg Foundation, Sweden; SERI, SNSF and Cantons of Bern and Geneva, Switzerland; NSTC, Taipei; TENMAK, T\"urkiye; STFC/UKRI, United Kingdom; DOE and NSF, United States of America.

Individual groups and members have received support from BCKDF, CANARIE, CRC and DRAC, Canada; CERN-CZ, FORTE and PRIMUS, Czech Republic; COST, ERC, ERDF, Horizon 2020, ICSC-NextGenerationEU and Marie Sk{\l}odowska-Curie Actions, European Union; Investissements d'Avenir Labex, Investissements d'Avenir Idex and ANR, France; DFG and AvH Foundation, Germany; Herakleitos, Thales and Aristeia programmes co-financed by EU-ESF and the Greek NSRF, Greece; BSF-NSF and MINERVA, Israel; NCN and NAWA, Poland; La Caixa Banking Foundation, CERCA Programme Generalitat de Catalunya and PROMETEO and GenT Programmes Generalitat Valenciana, Spain; G\"{o}ran Gustafssons Stiftelse, Sweden; The Royal Society and Leverhulme Trust, United Kingdom.

In addition, individual members wish to acknowledge support from Armenia: Yerevan Physics Institute (FAPERJ); CERN: European Organization for Nuclear Research (CERN PJAS); Chile: Agencia Nacional de Investigaci\'on y Desarrollo (FONDECYT 1230812, FONDECYT 1230987, FONDECYT 1240864); China: Chinese Ministry of Science and Technology (MOST-2023YFA1605700), National Natural Science Foundation of China (NSFC - 12175119, NSFC 12275265, NSFC-12075060); Czech Republic: Czech Science Foundation (GACR - 24-11373S), Ministry of Education Youth and Sports (FORTE CZ.02.01.01/00/22\_008/0004632), PRIMUS Research Programme (PRIMUS/21/SCI/017); EU: H2020 European Research Council (ERC - 101002463); European Union: European Research Council (ERC - 948254, ERC 101089007), Horizon 2020 Framework Programme (MUCCA - CHIST-ERA-19-XAI-00), European Union, Future Artificial Intelligence Research (FAIR-NextGenerationEU PE00000013), Italian Center for High Performance Computing, Big Data and Quantum Computing (ICSC, NextGenerationEU); France: Agence Nationale de la Recherche (ANR-20-CE31-0013, ANR-21-CE31-0013, ANR-21-CE31-0022, ANR-22-EDIR-0002), Investissements d'Avenir Labex (ANR-11-LABX-0012); Germany: Baden-Württemberg Stiftung (BW Stiftung-Postdoc Eliteprogramme), Deutsche Forschungsgemeinschaft (DFG - 469666862, DFG - CR 312/5-2); Italy: Istituto Nazionale di Fisica Nucleare (ICSC, NextGenerationEU), Ministero dell'Università e della Ricerca (PRIN - 20223N7F8K - PNRR M4.C2.1.1); Japan: Japan Society for the Promotion of Science (JSPS KAKENHI JP22H01227, JSPS KAKENHI JP22H04944, JSPS KAKENHI JP22KK0227, JSPS KAKENHI JP23KK0245); Netherlands: Netherlands Organisation for Scientific Research (NWO Veni 2020 - VI.Veni.202.179); Norway: Research Council of Norway (RCN-314472); Poland: Ministry of Science and Higher Education (IDUB AGH, POB8, D4 no 9722), Polish National Agency for Academic Exchange (PPN/PPO/2020/1/00002/U/00001), Polish National Science Centre (NCN 2021/42/E/ST2/00350, NCN OPUS nr 2022/47/B/ST2/03059, NCN UMO-2019/34/E/ST2/00393, UMO-2020/37/B/ST2/01043, UMO-2021/40/C/ST2/00187, UMO-2022/47/O/ST2/00148, UMO-2023/49/B/ST2/04085); Slovenia: Slovenian Research Agency (ARIS grant J1-3010); Spain: Generalitat Valenciana (Artemisa, FEDER, IDIFEDER/2018/048), Ministry of Science and Innovation (MCIN \& NextGenEU PCI2022-135018-2, MICIN \& FEDER PID2021-125273NB, RYC2019-028510-I, RYC2020-030254-I, RYC2021-031273-I, RYC2022-038164-I), PROMETEO and GenT Programmes Generalitat Valenciana (CIDEGENT/2019/027); Sweden: Swedish Research Council (Swedish Research Council 2023-04654, VR 2018-00482, VR 2022-03845, VR 2022-04683, VR 2023-03403, VR grant 2021-03651), Knut and Alice Wallenberg Foundation (KAW 2018.0157, KAW 2018.0458, KAW 2019.0447, KAW 2022.0358); Switzerland: Swiss National Science Foundation (SNSF - PCEFP2\_194658); United Kingdom: Leverhulme Trust (Leverhulme Trust RPG-2020-004), Royal Society (NIF-R1-231091); United States of America: U.S. Department of Energy (ECA DE-AC02-76SF00515), Neubauer Family Foundation.

%
%

% End of text imported from the .//acknowledgements/Acknowledgements.tex input file

%
%
%
%
%

%
%
%
%
%
%
%
%
%
%
%
%
%
%
%
%
%
%
%
%
%
%
%
%
%
%
%
%
%
%
%
%
%
%
%
%
%
%
%
%
%
%
%
%
%
%
%
%
%
%
%
%

%
%

%
%
%
%
%
%

%
%

%
%
%
\printbibliography
\clearpage
\input{atlas_authlist}
%

%
%

%

%
%
%
%

\end{document}

%% file: atlas_authlist.tex
% ATLAS Collaboration author list
% Reference date of SUSY-2018-37 is 2024-04-19
% Author list last updated on date 19-SEP-24
% Data extracted on 19-Sep-2024 for paper reference SUSY-2018-37
% at 10:11am
 
\begin{flushleft}
\hypersetup{urlcolor=black}
{\Large The ATLAS Collaboration}

\bigskip

\AtlasOrcid[0000-0002-6665-4934]{G.~Aad}$^\textrm{\scriptsize 104}$,
\AtlasOrcid[0000-0001-7616-1554]{E.~Aakvaag}$^\textrm{\scriptsize 17}$,
\AtlasOrcid[0000-0002-5888-2734]{B.~Abbott}$^\textrm{\scriptsize 123}$,
\AtlasOrcid[0000-0002-0287-5869]{S.~Abdelhameed}$^\textrm{\scriptsize 119a}$,
\AtlasOrcid[0000-0002-1002-1652]{K.~Abeling}$^\textrm{\scriptsize 56}$,
\AtlasOrcid[0000-0001-5763-2760]{N.J.~Abicht}$^\textrm{\scriptsize 50}$,
\AtlasOrcid[0000-0002-8496-9294]{S.H.~Abidi}$^\textrm{\scriptsize 30}$,
\AtlasOrcid[0009-0003-6578-220X]{M.~Aboelela}$^\textrm{\scriptsize 45}$,
\AtlasOrcid[0000-0002-9987-2292]{A.~Aboulhorma}$^\textrm{\scriptsize 36e}$,
\AtlasOrcid[0000-0001-5329-6640]{H.~Abramowicz}$^\textrm{\scriptsize 154}$
% ...
\bigskip
\\

% ...
$^{17}$Department for Physics and Technology, University of Bergen, Bergen; Norway.\\
% ...
$^{30}$Physics Department, Brookhaven National Laboratory, Upton NY; United States of America.\\
% ...
$^{36}$$^{(a)}$Facult\'e des Sciences Ain Chock, Universit\'e Hassan II de Casablanca;$^{(b)}$Facult\'{e} des Sciences, Universit\'{e} Ibn-Tofail, K\'{e}nitra;$^{(c)}$Facult\'e des Sciences Semlalia, Universit\'e Cadi Ayyad, LPHEA-Marrakech;$^{(d)}$LPMR, Facult\'e des Sciences, Universit\'e Mohamed Premier, Oujda;$^{(e)}$Facult\'e des sciences, Universit\'e Mohammed V, Rabat;$^{(f)}$Institute of Applied Physics, Mohammed VI Polytechnic University, Ben Guerir; Morocco.\\
% ...
$^{45}$Physics Department, Southern Methodist University, Dallas TX; United States of America.\\
% ...
$^{50}$Fakult\"{a}t Physik , Technische Universit{\"a}t Dortmund, Dortmund; Germany.\\
% ...
$^{56}$II. Physikalisches Institut, Georg-August-Universit\"{a}t G\"ottingen, G\"ottingen; Germany.\\
% ...
$^{104}$CPPM, Aix-Marseille Universit\'e, CNRS/IN2P3, Marseille; France.\\
% ...
$^{119}$$^{(a)}$New York University Abu Dhabi, Abu Dhabi;$^{(b)}$United Arab Emirates University, Al Ain; United Arab Emirates.\\
% ...
$^{123}$Homer L. Dodge Department of Physics and Astronomy, University of Oklahoma, Norman OK; United States of America.\\
% ...
$^{154}$Raymond and Beverly Sackler School of Physics and Astronomy, Tel Aviv University, Tel Aviv; Israel.\\
% ...
$^{a}$ Also Affiliated with an institute covered by a cooperation agreement with CERN.\\
$^{b}$ Also at An-Najah National University, Nablus; Palestine.\\
$^{c}$ Also at Borough of Manhattan Community College, City University of New York, New York NY; United States of America.\\
$^{d}$ Also at Center for Interdisciplinary Research and Innovation (CIRI-AUTH), Thessaloniki; Greece.\\
$^{e}$ Also at Centro Studi e Ricerche Enrico Fermi; Italy.\\
$^{f}$ Also at CERN, Geneva; Switzerland.\\
% ...

\end{flushleft}

% Created with Glance <Atlas.Glance@cern.ch>